\documentstyle[11pt]{article}

\newlength{\minitwocolumn}
\makeatletter
\@addtoreset{equation}{section}
\makeatother

\newtheorem{thm}{Theorem}[section]
\newtheorem{prop}[thm]{Proposition}
\newtheorem{lem}[thm]{Lemma}
\newtheorem{cor}[thm]{Corollary}

\begin{document}

\begin{flushright}
Univ. Melbourne \\
preprint No.25 \\
hep-th/9407191 \\
July 1994
\end{flushright}

\vspace{36pt}

\begin{center}
\begin{Large}
Polynomial identities of the Rogers--Ramanujan type

\vspace{24pt}

Omar Foda and Yas-Hiro Quano\raisebox{2mm}{$\star$}

\vspace{24pt}
{\it Department of Mathematics, University of Melbourne \\
 \it Parkville, Victoria 3052, Australia}
\end{Large}
\vspace{24pt}

\underline{ABSTRACT}

\end{center}

Presented are polynomial identities which imply
generalizations of Euler and Rogers--Ramanujan identities.
Both sides of the identities can be interpreted
as generating functions of certain restricted partitions.
We prove the identities by establishing
a graphical one-to-one correspondence
between those two kinds of restricted partitions.

\vspace{24pt}

\vfill
\hrule

\vskip 3mm
\begin{small}

\noindent\raisebox{2mm}{$\star$} Supported by
the Australian Research Council.

\end{small}

\newpage

\section{Introduction}

The Rogers--Ramanujan identities appear in combinatorial
problems of number theory \cite{Num}, lattice
statistical mechanics \cite{ESM},
and identities among Virasoro characters \cite{FNO,NRT,KNS,SB1}.
The analytic form of the Rogers--Ramanujan identities are
stated as follows \cite{Andbk}:
\begin{equation}
\displaystyle\prod_{n=1 \atop n \equiv \pm (j+1) ({\rm mod }5)}^{\infty}
(1-q^n )^{-1}
=
\displaystyle\sum_{r=0}^{\infty} \frac{q^{r^2 +jr}}{(q)_r}, ~~~~
j=0, 1,
\label{eqn:R-R}
\end{equation}
where $|q| <1$ and
\begin{equation}
(a)_n \equiv (a; q)_n =\frac{(a; q)_{\infty}}{(aq^n ;q)_{\infty}}, ~~~~
(a)_{\infty} \equiv (a; q)_{\infty} = \prod_{m=0}^{\infty}(1-aq^{m}).
\label{eqn:(a)}
\end{equation}
It is natural to define for positive integer $n$
$$
\frac{1}{(q)_{-n}} =\frac{(q^{1-n})_{\infty}}{(q)_{\infty}} =0.
$$
Consequently, if we introduce the following symbol
for a non negative integer $N$ and an integer $M$
$$
\left[ \begin{array}{c}
N \\ M \end{array}
\right]_q =
\displaystyle\frac{(q)_N}{(q)_M (q)_{N-M}},
$$
then it becomes polynomial when $0\leq M \leq N$, otherwise
it vanishes.

Gordon's generalization of
the Rogers--Ramanujan identities has the following
analytic form \cite{Andbk} for $|q|<1$ and $1\leq i \leq k$
\begin{equation}
\prod_{n=1 \atop n \not\equiv 0, \pm i {\rm (mod }2k+1)}^{\infty}
(1-q^n )^{-1}
=
\displaystyle\sum_{n_1 \geq \cdots \geq n_{k-1} \geq 0}
\frac{q^{n_1^2 +\cdots + n_{k-1}^2 +n_i +\cdots +n_{k-1}}}
     {(q)_{n_1 -n_2 } \cdots (q)_{n_{k-2}-n_{k-1}} (q)_{n_{k-1}}}.
\label{eqn:Gor}
\end{equation}
Thanks to Jacobi's triple product formula we can
recast (\ref{eqn:Gor}) as
\begin{equation}
\frac{1}{(q)_{\infty}}
\sum_{r=-\infty}^{\infty} (-1)^r q^{r((2k+1)r+2k-2i+1)/2}
=
\displaystyle\sum_{n_1 \geq \cdots \geq n_{k-1} \geq 0}
\frac{q^{n_1^2 +\cdots + n_{k-1}^2 +n_i +\cdots +n_{k-1}}}
     {(q)_{n_1 -n_2 } \cdots (q)_{n_{k-2}-n_{k-1}} (q)_{n_{k-1}}}.
\label{eqn:Gordon}
\end{equation}

In the process of a trial to prove several conjectures
obtained in \cite{SB1}, we encounter the following
polynomial identities:

\begin{thm}
Let $n, k, i$ be fixed non negative integers such that
$1 \leq i \leq k$. Then the following polynomial identity holds.
\begin{equation}
\begin{array}{cl}
&
\displaystyle\sum_{r=-\infty}^{\infty} (-1)^{r}
q^{r((2k+1)r+2k-2i+1)/2}
\left[ \begin{array}{c} n \\
\left[ \frac{n-k+i-(2k+1)r}{2} \right]
\end{array} \right]_q \\
= &
\displaystyle\sum_{
           {n_1 \geq \cdots \geq n_{k-1} \geq n_k =0}
     \atop {2(n_1 + \cdots + n_{k-1}) \leq n-k+i }}
q^{n_1^2 + \cdots + n_{k-1}^2 +n_{i} + \cdots + n_{k-1}} \times \\
\times &
\displaystyle\prod_{j=1}^{k-1}
\left[ \begin{array}{c}
n-2(n_1 + \cdots + n_{j-1})-n_j -n_{j+1} -\alpha^{(k)}_{ij} \\
n_j -n_{j+1}
\end{array} \right]_q .
\end{array}
\label{eqn:truncated}
\end{equation}
Here $[x]$ denotes the largest integer part of $x$, and
$\alpha^{(k)}_{ij}$ is the $(i,j)$th entry of the following
$k \times (k-1)$ matrix
\begin{equation}
A^{(k)}=
\left( \begin{array}{cccc}
1 & 2 & \cdots & k-1 \\
0 & 1 & \cdots & k-2 \\
\vdots & \ddots & \ddots & \vdots \\
0 & \cdots & 0 & 1 \\
0 & \cdots & 0 & 0
\end{array} \right).
\label{eqn:alpha}
\end{equation}
\label{thm:poly1}
\end{thm}

Note that both sides of
(\ref{eqn:truncated})
are
polynomials because all but finitely many terms vanish
in the sums and all the nonzero terms are polynomials.
The first special cases $k=1, 2$ of Theorem
\ref{thm:poly1} are due to Andrews \cite[Ch.9, Ex.4]{Andbk}.
Thus Theorem \ref{thm:poly1}
is a generalization of Andrews' polynomial identities.
You can reproduce (\ref{eqn:Gordon})
by taking the limit $n\rightarrow \infty$.

{\it Comments on historical matter.}
The case $k=2$ in \cite[Ch.9, Ex.4]{Andbk} reads
as follows:
\begin{equation}
\begin{array}{cl}
&
\displaystyle\sum_{r=-\infty}^{\infty} (-1)^{r}
q^{r(5r+1)/2-2ar}
\left[ \begin{array}{c} n \\
\left[ \frac{n-5r}{2} \right] + a
\end{array} \right]_q \\
= &
\displaystyle\sum_{j=0 \atop 2j \leq n-a }
q^{j^2 +aj }
\left[ \begin{array}{c}
n-j -a \\
j
\end{array} \right]_q ,
\end{array}
\label{eqn:API}
\end{equation}
where $a=0,1$\footnote{
Here we replace $n+1$ by $n$.
The $a=0$ and $a=1$ correspond to $k=i=2$
and $k=2, i=1$, respectively.
The LHS of (\ref{eqn:API}) for $a=1$ looks
different from ours for $k=2, i=1$,
but they actually coincide. }.
Proving $q$-series identities in terms of
polynomials with a finite parameter $n$
was initiated by Schur \cite{Sch},
who studied the LHS of (\ref{eqn:API})
in order to prove the Rogers--Ramanujan identities;
whereas the RHS of (\ref{eqn:API})
was found by MacMahon \cite{Mac}.
As an identity, (\ref{eqn:API})
appeared for the first time in \cite{And1}.

There exists another combinatorial identity by Euler
\cite[\S 1, Prob.23]{LL}
\begin{equation}
\prod_{n=1 \atop n \equiv 1-j ({\rm mod }2)}^{\infty} (1+q^n )
=
\displaystyle\sum_{r=0}^{\infty} \frac{q^{r^2 +jr}}{(q)_r (-q)_r }=
\displaystyle\sum_{r=0}^{\infty} \frac{q^{r^2 +jr}}{(q^2 ; q^2 )_r },
{}~~~~ j=0, 1.
\label{eqn:Euler}
\end{equation}
A generalization of (\ref{eqn:Euler})
is given by \cite{Bre2}
\begin{equation}
\frac{1}{(q)_{\infty}}
\prod_{n=1 \atop n \equiv 0, \pm i ({\rm mod }2k)} (1-q^n )
=
\displaystyle\sum_{n_1 \geq \cdots \geq n_{k-1} \geq 0}
\frac{q^{n_1^2 +\cdots + n_{k-1}^2 +n_{i} +\cdots +n_{k-1}}}
     {(q)_{n_1 -n_2 } \cdots (q)_{n_{k-2}-n_{k-1}}
      (q^2 ; q^2 )_{n_{k-1}}},
\label{eqn:GE}
\end{equation}
where $1 \leq i \leq k$.
It can be also recasted as
\begin{equation}
\frac{1}{(q)_{\infty}}
\sum_{r=-\infty}^{\infty} (-1)^r q^{r(kr+k-i)}
=
\displaystyle\sum_{n_1 \geq \cdots \geq n_{k-1} \geq 0}
\frac{q^{n_1^2 +\cdots + n_{k-1}^2 +n_{i} +\cdots +n_{k-1}}}
     {(q)_{n_1 -n_2 } \cdots (q)_{n_{k-2}-n_{k-1}}
      (q^2 ; q^2 )_{n_{k-1}}},
\label{eqn:GEuler}
\end{equation}
where $1 \leq i \leq k$.

It may be useful to express (\ref{eqn:Gordon})
and (\ref{eqn:GEuler}) in a unified form.
Let $L\geq 3$, $k=[L/2]$, and $1 \leq i \leq k$.
Then the following holds:
\begin{equation}
\frac{1}{(q)_{\infty}}
\sum_{r=-\infty}^{\infty} (-1)^r q^{r(Lr+L-2i)/2}
=
\displaystyle\sum_{n_1 \geq \cdots \geq n_{k-1} \geq 0}
\frac{q^{n_1^2 +\cdots + n_{k-1}^2 +n_{i} +\cdots +n_{k-1}}}
     {(q)_{n_1 -n_2 } \cdots (q)_{n_{k-2}-n_{k-1}}
      (q^a ; q^a )_{n_{k-1}}},
\end{equation}
where $a=1$ (resp. $a=2$) when $L$ is odd (resp. even).

We would like to also present
the following polynomial identity
which reduces to (\ref{eqn:GEuler})
in the limit $n \rightarrow \infty$:

\begin{thm}
Let $n, k, i$ be fixed non negative integers such that
$1 \leq i \leq k$ and $k\geq 2$.
Then the following polynomial identity holds.
\begin{equation}
\begin{array}{cl}
&
\displaystyle\sum_{r=-\infty}^{\infty} (-1)^{r}
q^{r(kr+k-i)}
\left[ \begin{array}{c} 2n+k-i \\
       n-kr
\end{array} \right]_q \\
= &
\displaystyle\sum_{
           {n_1 \geq \cdots \geq n_{k-1} \geq 0}
     \atop {n_1 + \cdots + n_{k-1} \leq n}}
q^{n_1^2 + \cdots + n_{k-1}^2 +n_{i} + \cdots + n_{k-1}} \times \\
\times &
\displaystyle\prod_{j=1}^{k-2}
\left[ \begin{array}{c}
2n-2(n_1 + \cdots + n_{j-1})-n_j -n_{j+1}
+\beta^{(k)}_{ij} \\
n_j -n_{j+1}
\end{array} \right]_q \times \\
\times &
\left[ \begin{array}{c}
n-(n_1 + \cdots + n_{k-2}) \\
n_{k-1}
\end{array} \right]_{q^2},
\end{array}
\label{eqn:finite}
\end{equation}
where $\beta^{(k)}_{ij}$ is the $(i,j)$th entry of the following
$k \times (k-2)$ matrix
\begin{equation}
B^{(k)}=
\left( \begin{array}{ccccc}
k-2 & k-3 & k-4 & \cdots & 1 \\
k-2 & k-3 & k-4 & \cdots & 1 \\
\vdots & & & & \vdots \\
3 & 2 & 1 & \cdots & 1 \\
2 & 1 & 1 & \cdots & 1 \\
1 & 1 & 1 & \cdots & 1 \\
0 & 0 & 0 & \cdots & 0
\end{array}
\right).
\label{eqn:beta}
\end{equation}
\label{thm:poly2}
\end{thm}

Note that both sides of (\ref{eqn:finite}) are again
polynomials
from the same reason as for (\ref{eqn:truncated}).

Now we would like to mention some related works on
Bose--Fermi correspondence of Virasoro characters.
As was discussed in \cite{SB1},
(\ref{eqn:Gordon}) is the simplest example
of $q$-series identities between bosonic and
fermionic representations (LHS and RHS, respectively)
of Virasoro characters.
In \cite{KR,KRV} character identities of the types
(\ref{eqn:Gordon}) and (\ref{eqn:GEuler})
were considered on the basis
of path space representations.
Theorems \ref{thm:poly1} and \ref{thm:poly2}
exhibit "finitized"\footnote{
This terminology is due to Melzer \cite{Mel1,Mel2}. } forms
of (\ref{eqn:Gordon}) and (\ref{eqn:GEuler}),
respectively. In \cite{Mel1}
polynomial identities
for characters
$\chi^{(\nu , \nu +1)}_{r,s}(q)$ of unitary minimal models
${\cal M}(\nu , \nu +1)$
were conjectured (and proved for $\nu =3,4$).
It was Berkovich \cite{Ber}
who proved these conjectures
for arbitrary $\nu $ and $s=1$.

The rest of this paper is organized as follows.
In section 2 we introduce a certain restricted partition
function
and we evaluate the associated
generating function by sieve technique
\cite{Andbk,Bre}. We thus obtain
the LHS's of (\ref{eqn:truncated})
and (\ref{eqn:finite}).
In section 3 we introduce another kind of restricted partition
functions
whose generating functions give
the RHS's of (\ref{eqn:truncated}) and (\ref{eqn:finite}).
We also present the main Propositions of this paper,
which shall be proved by the induction
with respect to $(k,i)$.
In section 4 we prove the first special cases $k=2$.
In section 5
we prove the main Propositions
by establishing a graphical one-to-one correspondence
between two kinds of partitions introduced in sections 2 and 3.
In section 6 we give discussion and remarks.

{}~

\section{Interpretation of LHS}

Let us begin by fixing several terms \cite{Andbk,Bre}.
Let the {\it rank} of a partition be the largest part minus
the number of parts.
For instance, the rank of
the partition $18=7+4+3+2+2$
is $7-5=2$.
In general, let $N=a_1 + \cdots + a_s$ be a partition of
$N$ such that $a_1 \geq \cdots \geq a_s$. Then
we construct the {\it Ferrers graph} of the partition
by putting $a_l$ dots on the $l$th row,
starting from the left.
Fig.\ref{fig:18=}
represents the Ferrers graph of the above partition.

\begin{minipage}[t]{\minitwocolumn}
\begin{center}
\ \vspace{2ex}\\
\setlength{\unitlength}{0.5mm}
\begin{picture}(50,35)
\put(-8,-5){\begin{picture}(101,0)
\put(4.5,4){\circle*{3}}
\put(4.5,14){\circle*{3}}
\put(4.5,24){\circle*{3}}
\put(4.5,34){\circle*{3}}
\put(4.5,44){\circle*{3}}
\put(14.5,4){\circle*{3}}
\put(14.5,14){\circle*{3}}
\put(14.5,24){\circle*{3}}
\put(14.5,34){\circle*{3}}
\put(14.5,44){\circle*{3}}
\put(24.5,24){\circle*{3}}
\put(24.5,34){\circle*{3}}
\put(24.5,44){\circle*{3}}
\put(34.5,34){\circle*{3}}
\put(34.5,44){\circle*{3}}
\put(44.5,44){\circle*{3}}
\put(54.5,44){\circle*{3}}
\put(64.5,44){\circle*{3}}
\end{picture}
}
\end{picture}
\vspace{2ex} \\
\refstepcounter{figure}
Fig.\thefigure : An example of partition of $18$.
\addcontentsline{lof}{figure}%
{\protect\numberline{\thefigure}{An example of partition of $18$.}}
\label{fig:18=}
\end{center}
\end{minipage}

{}~

The {\it subgraph} of a Ferrers graph
is a portion of
the Ferrers graph which lies below a given row
and to the right of a given column.
The $l$th {\it proper subgraph} is the subgraph
lying below the $l$th row and to the right of
the $l$th column.

The $l$th {\it right angle} refers to
a portion of the $(l-1)$st proper subgraph
minus the $l$th proper subgraph.
In the above example there are
three right angles.

The {\it length} of the $l$th row (resp. column)
refers to the number of dots on the $l$th
row (resp. column).
The {\it direction} parallel to rows (resp. column)
is horizontal (resp. vertical).
Let $\mu_l $ and $\nu_l $ be the lengths
of the $l$th row and column, respectively.
The $l$th {\it longer side} (resp. {\it shorter side})
{\it relative to $c$} refers to the $l$th row (resp. column)
if $\mu_l -\nu_l >c+1$.
The $l$th {\it longer side} (resp. {\it shorter side})
{\it relative to $c$} refers to the $l$th column (resp. row)
if $\mu_l -\nu_l <c$.
For the above example,
the length and the direction of the second longer side
relative to $1$ is $5$ and vertical, respectively.

Let $r_l (\pi)$ denote the $l$th {\it successive rank}
of a partition $\pi$,
the rank of the $(l-1)$st proper subgraph of
the corresponding Ferrers graph.
For the above partition $\pi $ we have
$r_1 (\pi )=2, r_2 (\pi )=-1, r_3 (\pi )=0$.
The $l$th successive rank is larger than $c+1$ (resp. less than $c$)
if and only if
the $l$th row is the $l$th longer (resp. shorter) side
relative to $c$.

For a given partition $\pi$
and positive integers $a,b$,
let $\lambda $ be the largest
integer for which there exists a sequence
$l_1 < \cdots < l_{\lambda}$ such that
$r_{l_1 }(\pi ) \geq a-1, r_{l_2 }(\pi )\leq -b+1,
r_{l_3 }(\pi ) \geq a-1, r_{l_4 }(\pi )\leq -b+1$,
and so on.
Then $\pi$ has an {\it $(a,b)$-positive oscillation
of length $\lambda$}.
For a given partition $\pi$
let $\lambda $ be the largest integer
for which there exists a sequence
$l_1 < \cdots < l_{\lambda}$ such that
$r_{l_1 }(\pi )\leq -b+1, r_{l_2 }(\pi ) \geq a-1,
r_{l_3 }(\pi )\leq -b+1, r_{l_4 }(\pi ) \geq a-1$,
and so on. Then $\pi $
has an {\it $(a,b)$-negative oscillation
of length $\lambda $}.

In what follows we will often consider partitions into
at most $\nu $ parts,
and with the largest part at most $\mu $.
A partition has the {\it maximal size} $(\mu , \nu)$
if each part does not exceed $\mu $, and
the number of parts does not
exceed $\nu$.

Let $p_{a,b}(\mu , \nu ; \lambda ; N)$
(resp. $m_{a,b}(\mu , \nu ; \lambda ; N)$)
stand for the number of partitions of $N$ with the
maximal size $(\mu , \nu )$, and with $(a,b)$-positive (resp. negative)
oscillation of length at least $\lambda$.
To such types partition functions we associate
the generating functions
$$
\begin{array}{rcl}
P_{a,b}(\mu , \nu ; \lambda ; q) & = &
\displaystyle\sum_{N\geq 0} p_{a,b}(\mu , \nu ; \lambda ; N) q^N , \\
M_{a,b}(\mu , \nu ; \lambda ; q) & = &
\displaystyle\sum_{N\geq 0} m_{a,b}(\mu , \nu ; \lambda ; N) q^N .
\end{array}
$$

The {\it main diagonal line} refers to the
diagonal line starting from the
top-left dot of the Ferrers graph.
For a given Ferrers graph,
we can obtain another Ferrers graph by reflecting
the original one with respect to the main diagonal line.
Let us call the graph thus
obtained the {\it dual Ferrers graph}
of the original one.
Note that the dual Ferrers graph of
any partition counted by $p_{a,b}(\mu , \nu ; \lambda ; N)$
gives a partition counted by
$m_{b,a}(\nu , \mu ; \lambda ; N)$,
and vice versa. Hence we obtain
\begin{equation}
\begin{array}{rcl}
p_{a,b}(\mu , \nu ; \lambda ; N) & = &
m_{b,a}(\nu , \mu ; \lambda ; N), \\
P_{a,b}(\mu , \nu ; \lambda ; q) & = &
M_{b,a}(\nu , \mu ; \lambda ; q).
\label{eqn:pm-sym}
\end{array}
\end{equation}

We cite a number of results
from \cite{Andbk} in extended forms.
You will know that those results
permit the following extension by reexamining
proofs given in \cite{Andbk}.

\begin{lem}
Let $p(\mu , \nu ; N)$ be the number of partitions of $N$
with the maximal size $(\mu ,\nu )$.
Then the generating function associated with this
partition function is given as follows:
\begin{equation}
P(\mu, \nu ; q)\equiv \sum_{N \geq 0} p(\mu , \nu ; N) q^N
=
\left[ \begin{array}{c}
\mu + \nu \\ \mu \end{array}
\right]_q
=\left[ \begin{array}{c}
\mu + \nu \\ \nu \end{array}
\right]_q .
\label{eqn:gfp}
\end{equation}
\label{lem:gfp}
\end{lem}

Since all partitions have $(a,b)$-positive and negative
oscillation of length more than or equal to $0$, both
$P_{a,b}(\mu , \nu ; 0 ; q)$
and $M_{a,b}(\mu , \nu ; 0 ; q)$ coincide with
$P(\mu, \nu ; q)$.
Consequently we obtain

\begin{lem}
\begin{equation}
P_{a,b}(\mu , \nu ; 0 ; q)=
M_{a,b}(\mu , \nu ; 0 ; q)=
\left[ \begin{array}{c}
\mu + \nu \\ \mu \end{array}
\right]_q
=\left[ \begin{array}{c}
\mu + \nu \\ \nu \end{array}
\right]_q .
\label{eqn:ini}
\end{equation}
\label{lem:ini}
\end{lem}

The definitions immediately derive the following lemma:

\begin{lem} For $\lambda \geq 1$,
the following hold:
\begin{equation}
P_{a,b}(0 , \nu ; \lambda ; q)=
P_{a,b}(\mu , 0 ; \lambda ; q)=
M_{a,b}(0 , \nu ; \lambda ; q)=
M_{a,b}(\mu , 0 ; \lambda ; q)= 0.
\label{eqn:ini2}
\end{equation}
\label{lem:ini2}
\end{lem}

\begin{lem}
The following recursion relation holds for $\lambda \geq 1$:
\begin{equation}
\begin{array}{cl}
& p_{a,b}(\mu , \nu ; \lambda ; N)-
p_{a,b}(\mu -1, \nu ; \lambda ; N)-
p_{a,b}(\mu , \nu -1; \lambda ; N)+
p_{a,b}(\mu -1, \nu -1; \lambda ; N) \\
= & \left\{ \begin{array}{ll}
p_{a,b}(\mu -1, \nu -1; \lambda ; N-\mu -\nu +1), &
\mbox{if $\mu -\nu \leq a-2$, } \\
m_{a,b}(\mu -1, \nu -1; \lambda -1; N-\mu -\nu +1), &
\mbox{if $\mu -\nu \geq a-1$, }
\end{array} \right.
\end{array}
\label{eqn:recp}
\end{equation}
\begin{equation}
\begin{array}{ll}
& m_{a,b}(\mu , \nu ; \lambda ; N)-
m_{a,b}(\mu -1, \nu ; \lambda ; N)-
m_{a,b}(\mu , \nu -1; \lambda ; N)+
m_{a,b}(\mu -1, \nu -1; \lambda ; N) \\
= & \left\{ \begin{array}{ll}
p_{a,b}(\mu -1, \nu -1; \lambda -1; N-\mu -\nu +1), &
\mbox{if $\mu -\nu \leq -b+1$, } \\
m_{a,b}(\mu -1, \nu -1; \lambda ; N-\mu -\nu +1), &
\mbox{if $\mu -\nu \geq -b+2$, }
\end{array} \right.
\end{array}
\label{eqn:recm}
\end{equation}
\label{lem:rec}
\end{lem}

As a Corollary of Lemma \ref{lem:rec}, we have

\begin{cor}
\begin{equation}
\begin{array}{cl}
& P_{a,b}(\mu , \nu ; \lambda ; q)-
P_{a,b}(\mu -1, \nu ; \lambda ; q)-
P_{a,b}(\mu , \nu -1; \lambda ; q)+
P_{a,b}(\mu -1, \nu -1; \lambda ; q) \\
= & q^{\mu +\nu -1} \times \left\{ \begin{array}{ll}
P_{a,b}(\mu -1, \nu -1; \lambda ; q), &
\mbox{if $\mu -\nu \leq a-2$, } \\
M_{a,b}(\mu -1, \nu -1; \lambda -1; q), &
\mbox{if $\mu -\nu \geq a-1$, }
\end{array} \right.
\end{array}
\label{eqn:recP}
\end{equation}
\begin{equation}
\begin{array}{ll}
& M_{a,b}(\mu , \nu ; \lambda ; q)-
M_{a,b}(\mu -1, \nu ; \lambda ; q)-
M_{a,b}(\mu , \nu -1; \lambda ; q)+
M_{a,b}(\mu -1, \nu -1; \lambda ; q) \\
= & q^{\mu +\nu -1} \times \left\{ \begin{array}{ll}
P_{a,b}(\mu -1, \nu -1; \lambda -1; q), &
\mbox{if $\mu -\nu \leq -b+1$, } \\
M_{a,b}(\mu -1, \nu -1; \lambda ; q), &
\mbox{if $\mu -\nu \geq -b+2$, }
\end{array} \right.
\end{array}
\label{eqn:recM}
\end{equation}
\label{cor:rec}
\end{cor}

The initial conditions (\ref{eqn:ini}--\ref{eqn:ini2})
and recursion relations (\ref{eqn:recP}--\ref{eqn:recM})
uniquely determine
$P_{a,b}(\mu , \nu ; \lambda ; q)$ and
$M_{a,b}(\mu , \nu ; \lambda ; q)$.
When the set $(a,\mu , \nu )$
(resp. $(b,\mu , \nu )$) satisfies a certain
relation, $P_{a,b}(\mu , \nu ; \lambda ; q)$
(resp. $M_{a,b}(\mu , \nu ; \lambda ; q)$) can be
expressed in a simple form.

\begin{lem} For positive integers $a, b$, let
$\mu , \nu $ be non negative integers such that
$\mu -\nu \leq a-1$. Then
$P_{a,b}(\mu , \nu ; \lambda ; q)$ has
the following expressions:
\begin{equation}
\begin{array}{rcl}
P_{a,b}(\mu , \nu ; 2\lambda ; q) & = &
q^{\lambda (2(a+b)\lambda -a+b)}
\displaystyle \left[ \begin{array}{c}
\mu + \nu \\ \mu- (a+b)\lambda \end{array}
\right]_q , \\
P_{a,b}(\mu , \nu ; 2\lambda -1 ; q) & = &
q^{(2\lambda -1)((a+b)\lambda -b)}
\displaystyle \left[ \begin{array}{c}
\mu + \nu \\ \mu -(a+b) \lambda +b \end{array}
\right]_q .
\end{array}
\label{eqn:expP}
\end{equation}
For positive integers $a, b$, let
$\mu , \nu $ be non negative integers such that
$\mu -\nu \geq -b+1$. Then $M_{a,b}(\mu , \nu ; \lambda ; q)$
has the following expressions:
\begin{equation}
\begin{array}{rcl}
M_{a,b}(\mu , \nu ; 2\lambda ; q) & = &
q^{\lambda (2(a+b)\lambda +a-b)}
\displaystyle \left[ \begin{array}{c}
\mu + \nu \\ \nu -(a+b)\lambda \end{array}
\right]_q , \\
M_{a,b}(\mu , \nu ; 2\lambda -1 ; q) & = &
q^{(2\lambda -1)((a+b)\lambda -a)}
\displaystyle \left[ \begin{array}{c}
\mu + \nu \\ \nu -(a+b) \lambda +a \end{array}
\right]_q .
\end{array}
\label{eqn:expM}
\end{equation}
\label{lem:exp}
\end{lem}

{\it Remark. } From (\ref{eqn:pm-sym}),
one of (\ref{eqn:expP}) and (\ref{eqn:expM})
implies the other.

In order to state the last lemma we
cite from \cite{Andbk}, we introduce
another kind of partition function.
Let $Q_{a,b}(\mu , \nu ; N)$ be
the number of partitions $\pi$
of $N \geq 0$ with the maximal size
$(\mu , \nu )$, and $-b+2 \leq r_l (\pi ) \leq a-2$ for
any successive ranks of $\pi$.

\begin{lem} Three kinds of partition functions
$Q_{a,b}(\mu , \nu ; N)$, $p_{a,b}(\mu , \nu ; \lambda ; N)$ and
$m_{a,b}(\mu , \nu ; \lambda ; N)$ satisfy the following relation:
\begin{equation}
Q_{a,b}(\mu , \nu ; N)=
p_{a,b}(\mu , \nu ; 0 ; N) +
\sum_{\lambda =1}^{\infty} (-1)^{\lambda }
p_{a,b}(\mu , \nu ; \lambda ; N) +
\sum_{\lambda =1}^{\infty} (-1)^{\lambda }
m_{a,b}(\mu , \nu ; \lambda ; N).
\label{eqn:sieve}
\end{equation}
\label{lem:sieve}
\end{lem}

{\bf Proof. } This can be shown by using sieve technique.

The LHS of (\ref{eqn:sieve}) counts all partitions of $N$
with the maximal size $(\mu , \nu)$,
and with $(a,b)$-positive and negative oscillation of
length $0$. On the other hand, the first term in the RHS
refers to the number of partitions of $N$ with the
maximal size $(\mu , \nu )$. Subtract
the number of partitions
which have $(a,b)$-positive or negative oscillation of length
at least $1$ form the first term:
$$
p_{a,b}(\mu , \nu ; 0 ; N) -
p_{a,b}(\mu , \nu ; 1 ; N) -
m_{a,b}(\mu , \nu ; 1 ; N).
$$
In this way, however,
partitions which have both
$(a,b)$-positive and negative oscillation of length
at least $1$, are subtracted twice. Hence we have to add
the number of partitions
which have $(a,b)$-positive or negative oscillation of length
at least $2$:
$$
p_{a,b}(\mu , \nu ; 0 ; N) -
p_{a,b}(\mu , \nu ; 1 ; N) -
m_{a,b}(\mu , \nu ; 1 ; N) +
p_{a,b}(\mu , \nu ; 2 ; N) +
m_{a,b}(\mu , \nu ; 2 ; N).
$$
Note that
$
2\sum_{i =1}^{\lambda -1} (-1)^{i}
=0$ (resp. $-2$) when $\lambda $ is odd (resp. even).
Thus in general,
partitions which have $(a,b)$-positive or negative oscillation
of length at least $2\lambda -1$ should be subtracted
once more, and
those which have $(a,b)$-positive or negative oscillation
of length at least $2\lambda $ should be added once more.

Since the alternating sums in the RHS of (\ref{eqn:sieve})
are actually finite sums for fixed $\mu , \nu $ and $N$,
after repeating this procedure finitely many times,
we obtain (\ref{eqn:sieve}). ~~~~$\Box$

Let us introduce the generating function
of $Q_{a, b}(\mu , \nu ; N)$
as follows:
$$
{\cal Q}_{a,b}(\mu , \nu ; q) =
\sum_{N\geq 0} Q_{a,b}(\mu , \nu ; N) q^N .
$$
Now we wish to show that for
appropriate sets $(a, b, \mu , \nu)$,
${\cal Q}_{a, b}(\mu , \nu ; q)$ coincides with the LHS's of
(\ref{eqn:truncated}) and (\ref{eqn:finite}):

\begin{prop}
Let $a=2k+1-i, b=i$ and
$\mu =\left[ \frac{n+1+k-i }{2}\right]$,
$\nu =\left[ \frac{n-k+i}{2}\right]$,
where $n$ is a non negative integer.
Then ${\cal Q}_{a,b}(\mu , \nu ; q)$
gives the LHS of
(\ref{eqn:truncated}).
\label{prop:LHS1}
\end{prop}

{\bf Proof.}
First we notice that $a+b=n$. Furthermore,
$-b+1 \leq \mu -\nu \leq a-1$ holds because
$\mu -\nu =k-i$ when $n \equiv k-i$ mod $2$,
and $\mu -\nu =k-i+1$ when $n \not\equiv k-i$ mod $2$.
Hence from Lemmas \ref{lem:ini}, \ref{lem:exp}
and \ref{lem:sieve}
we have
$$
\begin{array}{cl}
& \displaystyle{\cal Q}_{a,b}(\mu , \nu ; N) \\
= &
\displaystyle\sum_{N\geq 0} p_{a,b}(\mu , \nu ; 0 ; N) q^N +
\sum_{N\geq 0} \sum_{\lambda =1}^{\infty} (-1)^{\lambda }
p_{a,b}(\mu , \nu ; \lambda ; N) q^N +
\sum_{N\geq 0} \sum_{\lambda =1}^{\infty} (-1)^{\lambda }
m_{a,b}(\mu , \nu ; \lambda ; N) q^N \\
= &
\displaystyle P_{a,b}(\mu , \nu ; 0 ; q)
+\sum_{\lambda =1}^{\infty} (-1)^{\lambda} P_{a,b}(\mu , \nu ; \lambda ; q)
+\sum_{\lambda =1}^{\infty} (-1)^{\lambda} M_{a,b}(\mu , \nu ; \lambda ; q) \\
= &
\displaystyle\left[ \begin{array}{c}
\mu + \nu \\ \nu \end{array}
\right]_q +
\sum_{\lambda =1}^{\infty}
q^{\lambda (2(a+b)\lambda -a+b)}
\displaystyle \left[ \begin{array}{c}
\mu + \nu \\ \mu -(a+b)\lambda \end{array}
\right]_q  \\
- &  \displaystyle\sum_{\lambda =1}^{\infty}
q^{(2\lambda -1)((a+b)\lambda -b)}
\left[ \begin{array}{c}
\mu + \nu \\ \mu -(a+b) \lambda +b \end{array}
\right]_q  \\
+ & \displaystyle\sum_{\lambda =1}^{\infty}
q^{\lambda(2(a+b)\lambda +a-b)}
\displaystyle \left[ \begin{array}{c}
\mu + \nu \\ \nu -(a+b)\lambda \end{array}
\right]_q - \sum_{\lambda =1}^{\infty}
q^{(2\lambda -1)((a+b)\lambda -a)}
\displaystyle \left[ \begin{array}{c}
\mu + \nu \\ \nu -(a+b) \lambda +a \end{array}
\right]_q \\
\end{array}
$$
$$
\begin{array}{cl}
= &
\displaystyle\sum_{\lambda =-\infty}^{\infty}
\left(
q^{\lambda (2(a+b)\lambda +a-b)}
\displaystyle \left[ \begin{array}{c}
\mu + \nu \\ \nu-(a+b)\lambda \end{array}
\right]_q -
q^{(2\lambda -1)((a+b)\lambda -b)}
\displaystyle \left[ \begin{array}{c}
\mu + \nu \\ \mu -(a+b) \lambda +b \end{array}
\right]_q \right) \\
= &
\displaystyle\sum_{r=-\infty}^{\infty} (-1)^{r}
q^{r((2k+1)r+2k-2i+1)/2}
\left[ \begin{array}{c} n \\
\left[ \frac{n-k+i-(2k+1)r}{2} \right]
\end{array} \right]_q .
\end{array}
$$
Thus the claim of this proposition was verified. ~~~~$\Box$

Now it is very easy to prove Theorem \ref{thm:poly1}
for $i=k=1$.
In this case
the LHS of (\ref{eqn:truncated})
is equal to ${\cal Q}_{2,1}([\frac{n+1}{2}], [\frac{n}{2}]; q)$.
On the other hand, $Q_{2,1}(\mu , \nu ; N)$
counts the number of partitions $\pi $ of $N$ with
the maximal size $(\mu , \nu )$, and
$1 \leq r_l (\pi ) \leq 0$ for all $l$.
Hence by the definition of the generating function, we have
${\cal Q}_{2,1}([\frac{n+1}{2}], [\frac{n}{2}]; q)=1$.
Thus we obtain
\begin{equation}
\sum_{r=-\infty}^{\infty} (-1)^{r}
q^{r(3r+1)/2}
\left[ \begin{array}{c} n \\
\left[ \frac{n-3r}{2} \right]
\end{array} \right]_q =1,
\label{eqn:1,1}
\end{equation}
which is nothing but Theorem \ref{thm:poly1}
for $i=k=1$.

The analogue of Proposition \ref{prop:LHS1}
for Theorem \ref{thm:poly2} is given as follows:

\begin{prop} Let $a=2k-i, b=i$ and
$\mu =n+k-i$, $\nu =n$,
where $n$ is a non negative integer.
Then
${\cal Q}_{a, b}(\mu , \nu ; q)$ gives the LHS of
(\ref{eqn:finite}).
\label{prop:LHS2}
\end{prop}

{\bf Proof. } First we notice that
$-b+1 \leq \mu -\nu \leq a-1$.
Hence from the same calculation
as in the proof of
Proposition \ref{prop:LHS1} we have
$$
\begin{array}{cl}
& \displaystyle{\cal Q}_{a,b}(\mu , \nu ; q) \\
= &
\displaystyle\sum_{\lambda =-\infty}^{\infty}
\left(
q^{\lambda (2(a+b)\lambda +a-b)}
\displaystyle \left[ \begin{array}{c}
\mu + \nu \\ \nu-(a+b)\lambda \end{array}
\right]_q -
q^{(2\lambda -1)((a+b)\lambda -b)}
\displaystyle \left[ \begin{array}{c}
\mu + \nu \\ \mu -(a+b) \lambda +b \end{array}
\right]_q \right) \\
= &
\displaystyle\sum_{r=-\infty}^{\infty} (-1)^{r}
q^{r(kr+k-i)}
\left[ \begin{array}{c} 2n+k-i \\
n-kr \end{array} \right]_q .
\end{array}
$$
Thus the claim of this proposition was verified. ~~~~$\Box$

{}~

\section{Main Propositions}

For a fixed set $(k,i)$ such that
$1 \leq i \leq k$
and a non negative integer $n$,
let $d(k,i;n;N)$ be the number of partitions of $N$
of the form
\begin{equation}
N=n_1^2 + \cdots + n_{k-1}^2 +n_i + \cdots + n_{k-1}
+\sum_{j=1}^{k-1} \sum_{p=1}^{n_j -n_{j+1}} a^{(j)}_{p},
\label{eqn:restd1}
\end{equation}
where
\begin{equation}
n_1 \geq n_2 \geq \cdots \geq n_{k-1}\geq n_k =0, ~~~~
2(n_1 + \cdots + n_{k-1}) \leq n-k+i,
\label{eqn:cond1}
\end{equation}
and
\begin{equation}
n-2(n_1 + \cdots + n_j )-
\alpha^{(k)}_{ij}
\geq a^{(j)}_1 \geq \cdots \geq a^{(j)}_{n_j -n_{j+1}}
\geq 0, ~~~~1 \leq j \leq k-1.
\label{eqn:fincond1}
\end{equation}
Since $\alpha^{(k)}_{i k-1}=k-i$,
(\ref{eqn:fincond1}) implies the second
condition of (\ref{eqn:cond1}).
We call (\ref{eqn:fincond1}) the finiteness condition
for partitions of the form (\ref{eqn:restd1}).

We notice that by taking into account Lemma \ref{lem:gfp},
the RHS of (\ref{eqn:truncated})
is the generating function of $d(k,i;n;N)$.
On the other hand, we showed
in the last section that
the LHS of (\ref{eqn:truncated}) coincides with
${\cal Q}_{2k+1-i,i}([\frac{n+1+k-i }{2}],
                     [\frac{n-k+i}{2}] ; q)$.
Thus what we should establish in order to show
Theorem \ref{thm:poly1} is that
there exists a one-to-one correspondence
between restricted partitions
counted by $d(k,i;n;N)$ and those counted by
$Q_{2k+1-i,i}([\frac{n+1+k-i }{2}],
              [\frac{n-k+i}{2}] ; N)$.
Actually, more than this is true:

\begin{prop} Fix a set $(k,i)$ such that
$1 \leq i \leq k$
and non negative integers $n, n_1, N$.
Let $d(k,i;n ; n_1 ; N)$
denote the number of partitions of $N$ of the form
(\ref{eqn:restd1}), and let

\noindent $Q_{2k+1-i,i}
([\frac{n+1+k-i }{2}], [\frac{n-k+i}{2}] ; n_1 ; N)$
denote the number of partitions
of $N$ with the maximal size
$([\frac{n+1+k-i }{2}], [\frac{n-k+i}{2}] )$,
with $n_1 $ right angles,
and all $n_1 $ successive ranks belong to
the interval $[-i+2,2k-i-1]$.
Then there exists a one-to-one correspondence
between restricted partitions
counted by $d(k,i;n; n_1 ; N)$ and those counted by
$Q_{2k+1-i,i}([\frac{n+1+k-i }{2}],[\frac{n-k+i}{2}];n_1 ; N)$.
\label{prop:k,i,1}
\end{prop}

Let us introduce another restricted partition.
For a fixed set $(k,i)$ such that
$k \geq 2, 1 \leq i \leq k$
and a non negative integer $n$,
let $\delta (k,i;n;N)$ be the number of partitions of $N$
of the form
\begin{equation}
N=n_1^2 + \cdots + n_{k-1}^2 +n_i + \cdots + n_{k-1}
+\sum_{j=1}^{k-2} \sum_{p=1}^{n_j -n_{j+1}}
a^{(j)}_{p} +2 \sum_{p=1}^{n_{k-1}} b_p ,
\label{eqn:restd2}
\end{equation}
where
\begin{equation}
n_1 \geq n_2 \geq \cdots \geq n_{k-1}\geq 0, ~~~~
n_1 + \cdots + n_{k-1} \leq n,
\label{eqn:cond2}
\end{equation}
and
\begin{equation}
\begin{array}{l}
2n-2(n_1 + \cdots + n_j )+
\beta^{(k)}_{ij} \geq a^{(j)}_1 \geq \cdots \geq a^{(j)}_{n_j -n_{j+1}}
\geq 0, ~~~~1 \leq j \leq k-2, \\
n-(n_1 + \cdots +n_{k-1}) \geq b_1 \geq \cdots \geq b_{n_{k-1}} \geq 0.
\end{array}
\label{eqn:fincond2}
\end{equation}
Note that the second condition of
(\ref{eqn:fincond2}) implies the second
one of (\ref{eqn:cond2}).
We call (\ref{eqn:fincond2}) the finiteness condition
for partitions of the form (\ref{eqn:restd2}).

By taking into account Lemma \ref{lem:gfp},
the RHS of (\ref{eqn:finite})
is the generating function of $\delta (k,i;n;N)$.
On the other hand, we showed
in the last section that
the LHS of (\ref{eqn:finite}) gives
${\cal Q}_{2k-i,i}(n+k-i , n ; q)$.
Thus what we should establish
in order to show Theorem \ref{thm:poly2} is that
there exists a one-to-one correspondence
between restricted partitions
counted by $\delta (k,i;n;N)$ and those counted by
$Q_{2k-i,i}(n+k-i , n ; N)$.
In this case, the following folds:

\begin{prop} Fix a set $(k,i)$ such that
$k \geq 2, 1 \leq i \leq k$
and non negative integers $n, n_1, N$.
Let $\delta (k,i;n ; n_1 ; N)$
denote the number of partitions of $N$ of the form
(\ref{eqn:restd2}), and let $Q_{2k-i,i}(n+k-i , n ; n_1 ; N)$
denote the number of partitions
of $N$ with the maximal size $(n+k-i, n)$,
with $n_1 $ right angles,
and all $n_1 $ successive ranks belong to
the interval $[-i+2,2k-i-2]$.
Then there exists a one-to-one correspondence
between restricted partitions
counted by $\delta (k,i;n; n_1 ; N)$ and those counted by
$Q_{2k-i,i}(n+k-i , n; n_1; N)$.
\label{prop:k,i,2}
\end{prop}

Propositions \ref{prop:k,i,1} and \ref{prop:k,i,2}
are the main Propositions of the present paper.
We prove these by induction
with respect to $(k,i)$ in section 4 and 5.
In section 4 we verify the first special cases $k=2$.
In section 5 we establish a graphical one-to-one
correspondence which implies the claims of
the main Propositions.
We notice that
a graphical correspondence presented below
can be obtained by
translating Burge's method \cite{Bur,AB,Bre4} into
language of partitions.

{}~

\section{The first special cases}

In this section we show the first special
cases.
The following is Proposition \ref{prop:k,i,1}
for $k=2, i=2$.

\begin{lem}
Fix non negative integers $n, n_1 , N$
such that $2n_1 \leq n$. Then
there exists a one-to-one correspondence
between a partition of $N$ of the form
\begin{equation}
N=n_1^2 +\sum_{p=1}^{n_{1}} a_p , ~~~~
n-2n_1 \geq a_1 \geq  \cdots \geq a_{n_1} \geq 0,
\label{eqn:k=2,i=2,1}
\end{equation}
and
a partition counted by $Q_{3,2}
([\frac{n+1}{2}], [\frac{n}{2}] ; n_1; N)$.
\label{lem:k=2,i=2,1}
\end{lem}

{\bf Proof. }
Let us recall that $Q_{3,2}([\frac{n+1}{2}],
[\frac{n}{2}] ; n_1 ; N)$
counts partitions of $N$ with the maximal size
$([\frac{n+1}{2}],[\frac{n}{2}])$,
with $n_1 $ right angles,
and all $n_1 $ successive ranks are equal to $0$
or $1$.
Note that the $p$th successive rank should be
equal to $0$ (resp. $1$) when the number of dots
of the $p$th right angle is odd (resp. even).
Thus we can construct a Ferrers graph of
partition counted by
$Q_{3,2}([\frac{n+1}{2}], [\frac{n}{2}] ; n_1 ; N)$
as follows.
If the number of dots of the $p$th right angle is odd,
we array
dots of the $p$th right angle
in a symmetric manner with respect to
the main diagonal line.
If the number of dots of $p$th right angle is even,
we array
dots on the the $p$th row one more than dots on the $p$th column.
Since the numbers of dots of
adjacent right angles differ by at least $2$,
the graph thus obtained always becomes a Ferrers graph.

Let us count the number of dots as follows.
First remove the $n_1 \times n_1 $
square in the top-left corner,
and count the number of dots on the $p$th row
to the right of the square plus
that on the $p$th column below the square.
In this way, we get a partition of $N$ into
$n_1^2 $ and at most $n_1 $ positive integers
less than or equal to
$n-2n_1 $.
Inversely, for any given partition of $N$
of the form (\ref{eqn:k=2,i=2,1}),
we can construct a Ferrers graph
with the maximal size $([\frac{n+1}{2}],[\frac{n}{2}])$,
and all $n_1 $ successive ranks are equal to $0$ or $1$,
by putting $n_1 \times n_1 $ dots, and
after that by
putting $[\frac{a_p+1}{2}]$ dots
on the $p$th row to the right of the square and
$[\frac{a_p }{2}]$
dots on the $p$th column below the square, respectively,
where $1 \leq p \leq n_1 $.
Thus the claim was established. ~~~~$\Box$

\begin{minipage}[t]{\minitwocolumn}
\begin{center}
\ \vspace{2ex}\\
\setlength{\unitlength}{0.5mm}
\begin{picture}(50,50)
\put(-8,-5){\begin{picture}(101,0)
\put(-1,29){\line(1,0){30}}
\put(-1,39){\line(1,0){30}}
\put(-1,49){\line(1,0){30}}
\put(-1,59){\line(1,0){30}}
\put(-1,29){\line(0,1){30}}
\put(9,29){\line(0,1){30}}
\put(19,29){\line(0,1){30}}
\put(29,29){\line(0,1){30}}
\put(4.5,4){\circle*{3}}
\put(4.5,14){\circle*{3}}
\put(4.5,24){\circle*{3}}
\put(4.5,34){\circle*{3}}
\put(4.5,44){\circle*{3}}
\put(4.5,54){\circle*{3}}
\put(14.5,24){\circle*{3}}
\put(14.5,34){\circle*{3}}
\put(14.5,44){\circle*{3}}
\put(14.5,54){\circle*{3}}
\put(24.5,34){\circle*{3}}
\put(24.5,44){\circle*{3}}
\put(24.5,54){\circle*{3}}
\put(34.5,34){\circle*{3}}
\put(34.5,44){\circle*{3}}
\put(34.5,54){\circle*{3}}
\put(44.5,44){\circle*{3}}
\put(44.5,54){\circle*{3}}
\put(54.5,54){\circle*{3}}
\end{picture}
}
\end{picture}
\vspace{2ex} \\
\refstepcounter{figure}
Fig.\thefigure : An example of partition counted by $Q_{3,2}(6,6:19)$.
\addcontentsline{lof}{figure}%
{\protect\numberline{\thefigure}
{An example of partition counted by $Q_{3,2}(6,6:19)$.}}
\label{fig:(3,2)}
\end{center}
\end{minipage}

{}~

The following is Proposition \ref{prop:k,i,1}
for $k=2, i=1$.

\begin{lem}
Fix non negative integers $n, n_1 , N$
such that $2n_1 \leq n-1$. Then
there exists a one-to-one correspondence
between a partition of $N$ of the form
\begin{equation}
N=n_1 (n_1 +1) +\sum_{p=1}^{n_{1}} a_p , ~~~~
n-2n_1 -1 \geq a_1 \geq  \cdots \geq a_{n_1} \geq 0,
\label{eqn:k=2,i=1,1}
\end{equation}
and
a partition counted by $Q_{4,1}
([\frac{n+2}{2}], [\frac{n-1}{2}] ; n_1; N)$.
\label{lem:k=2,i=1,1}
\end{lem}

{\bf Proof. }
Let us recall that all $n_1 $
successive ranks of partition counted by
$Q_{4,1}([\frac{n+2}{2}], [\frac{n-1}{2}] ; n_1 ; N)$
are equal to $1$ or $2$.
Hence the number of each right
angle is at least $2$.
In this case we can construct a Ferrers graph
of such a partition as follows. If the $p$th right angle
has $d_p -1$ ($\geq 2$) dots, we put
$[\frac{d_p +2}{2}]$ dots on the horizontal part
of the $p$th right angle including the dot on
the main diagonal line, and
$[\frac{d_p -1}{2}]$ dots on the vertical part
of the $p$th right angle including the dot
on the main diagonal line.
Now we can count the number of dots
by removing the $n_1 \times (n_1 +1)$
rectangle in the top-left corner,
and by counting the number of dots
on the $p$th row
to the right of the rectangle plus
that on the $p$th column below the rectangle.
Then we obtain a partition of $N$
into $n_1 (n_1 +1)$ and at most
$n_1 $ positive integers less than
or equal to $n-2n_1 -1$.
Thus the claim was verified.
{}~~~~$\Box$

\begin{minipage}[t]{\minitwocolumn}
\begin{center}
\ \vspace{2ex}\\
\setlength{\unitlength}{0.5mm}
\begin{picture}(50,50)
\put(-18,-5){\begin{picture}(101,0)
\put(-1,29){\line(1,0){40}}
\put(-1,39){\line(1,0){40}}
\put(-1,49){\line(1,0){40}}
\put(-1,59){\line(1,0){40}}
\put(-1,29){\line(0,1){30}}
\put(9,29){\line(0,1){30}}
\put(19,29){\line(0,1){30}}
\put(29,29){\line(0,1){30}}
\put(39,29){\line(0,1){30}}
\put(4.5,14){\circle*{3}}
\put(4.5,24){\circle*{3}}
\put(4.5,34){\circle*{3}}
\put(4.5,44){\circle*{3}}
\put(4.5,54){\circle*{3}}
\put(14.5,24){\circle*{3}}
\put(14.5,34){\circle*{3}}
\put(14.5,44){\circle*{3}}
\put(14.5,54){\circle*{3}}
\put(24.5,34){\circle*{3}}
\put(24.5,44){\circle*{3}}
\put(24.5,54){\circle*{3}}
\put(34.5,34){\circle*{3}}
\put(34.5,44){\circle*{3}}
\put(34.5,54){\circle*{3}}
\put(44.5,44){\circle*{3}}
\put(44.5,54){\circle*{3}}
\put(54.5,44){\circle*{3}}
\put(54.5,54){\circle*{3}}
\put(64.5,54){\circle*{3}}
\end{picture}
}
\end{picture}
\vspace{2ex} \\
\refstepcounter{figure}
Fig.\thefigure : An example of partition counted by $Q_{4,1}(7,6:20)$.
\addcontentsline{lof}{figure}%
{\protect\numberline{\thefigure}
{An example of partition counted by $Q_{4,1}(7,6:20)$.}}
\label{fig:(4,1)}
\end{center}
\end{minipage}

{}~

Next we consider the first special cases
of Proposition \ref{prop:k,i,2}.
The following is Proposition \ref{prop:k,i,2}
for $k=2, i=2$.

\begin{lem}
Fix non negative integers $n, n_1 , N$
such that $n_1 \leq n$. Then
there exists a one-to-one correspondence
between a partition of $N$ of the form
\begin{equation}
N=n_1^2 +2 \sum_{p=1}^{n_{1}} b_p , ~~~~
n-n_1 \geq b_1 \geq  \cdots \geq b_{n_1} \geq 0,
\label{eqn:k=2,i=2,2}
\end{equation}
and
a partition counted by $Q_{2,2}(n, n ; n_1; N)$.
\label{lem:k=2,i=2,2}
\end{lem}

{\bf Proof. }
Let us recall that $Q_{2,2}(n, n ; n_1 ; N)$
counts partitions of $N$ with the maximal size $(n,n)$,
with $n_1 $ right angles,
and all $n_1 $ successive ranks are equal to $0$.
Any Ferrers graph corresponding to such a partition
is symmetric with respect to the main diagonal line.
Let us count the number of dots as follows.
First remove the $n_1 \times n_1 $
square in the top-left corner,
and count the number of dots on
the $p$th row to the right of the square plus
that on the $p$th column below the square.
In this way, we get a partition of $N$ into
$n_1^2 $ and at most $n_1 $ even positive integers
less than or equal to
$2(n-n_1 )$.
Inversely, for any given partition of $N$
of the form (\ref{eqn:k=2,i=2,2}),
we can construct a symmetric Ferrers graph
by putting $n_1 \times n_1 $ dots, and
after that for $1 \leq p \leq n_1 $ by
putting $b_p$ dots on the $p$th row
to the right of the square and $b_p$ dots
on the $p$th column below the square, respectively.
Thus the claim was established. ~~~~$\Box$

\begin{minipage}[t]{\minitwocolumn}
\begin{center}
\ \vspace{2ex}\\
\setlength{\unitlength}{0.5mm}
\begin{picture}(50,50)
\put(-8,-5){\begin{picture}(101,0)
\put(-1,29){\line(1,0){30}}
\put(-1,39){\line(1,0){30}}
\put(-1,49){\line(1,0){30}}
\put(-1,59){\line(1,0){30}}
\put(-1,29){\line(0,1){30}}
\put(9,29){\line(0,1){30}}
\put(19,29){\line(0,1){30}}
\put(29,29){\line(0,1){30}}
\put(4.5,4){\circle*{3}}
\put(4.5,14){\circle*{3}}
\put(4.5,24){\circle*{3}}
\put(4.5,34){\circle*{3}}
\put(4.5,44){\circle*{3}}
\put(4.5,54){\circle*{3}}
\put(14.5,24){\circle*{3}}
\put(14.5,34){\circle*{3}}
\put(14.5,44){\circle*{3}}
\put(14.5,54){\circle*{3}}
\put(24.5,34){\circle*{3}}
\put(24.5,44){\circle*{3}}
\put(24.5,54){\circle*{3}}
\put(34.5,44){\circle*{3}}
\put(34.5,54){\circle*{3}}
\put(44.5,54){\circle*{3}}
\put(54.5,54){\circle*{3}}
\end{picture}
}
\end{picture}
\vspace{2ex} \\
\refstepcounter{figure}
Fig.\thefigure : An example of partition counted by $Q_{2,2}(6,6:17)$.
\addcontentsline{lof}{figure}%
{\protect\numberline{\thefigure}
{An example of partition counted by $Q_{2,2}(6,6:17)$.}}
\label{fig:(2,2)}
\end{center}
\end{minipage}

{}~

{\it Remark. } For a symmetric
Ferrers graph,
every right angle has odd number of dots.
Hence ${\cal Q}_{2,2}(n,n;q)$
gives the generating function of
partitions into distinct odd positive
integers less than or equal to $2n-1$:
$$
{\cal Q}_{2,2}(n,n;q)= \prod_{j=1}^{n} (1+q^{2j-1}).
$$

{}~

The following is Proposition \ref{prop:k,i,2}
for $k=2, i=1$.

\begin{lem}
Fix non negative integers $n, n_1 , N$
such that $n_1 \leq n$. Then
there exists a one-to-one correspondence
between a partition of $N$ of the form
\begin{equation}
N=n_1 (n_1 +1) +2 \sum_{p=1}^{n_{1}} b_p , ~~~~
n-n_1 \geq b_1 \geq  \cdots \geq b_{n_1} \geq 0,
\label{eqn:k=2,i=1,2}
\end{equation}
and
a partition counted by $Q_{3,1}(n+1, n ; n_1; N)$.
\label{lem:k=2,i=1,2}
\end{lem}

{\bf Proof. }
Let us recall that all the successive ranks
of a partition counted by $Q_{3,1}(n+1, n ; n_1 ; N)$
are equal to $1$.
Hence in this case,
we can count the number of dots
by removing the $n_1 \times (n_1 +1)$
rectangle in the top-left corner,
instead of the $n_1 \times n_1 $ square.
Thus the claim was verified.
{}~~~~$\Box$

\begin{minipage}[t]{\minitwocolumn}
\begin{center}
\ \vspace{2ex}\\
\setlength{\unitlength}{0.5mm}
\begin{picture}(50,50)
\put(-8,-5){\begin{picture}(101,0)
\put(-1,29){\line(1,0){40}}
\put(-1,39){\line(1,0){40}}
\put(-1,49){\line(1,0){40}}
\put(-1,59){\line(1,0){40}}
\put(-1,29){\line(0,1){30}}
\put(9,29){\line(0,1){30}}
\put(19,29){\line(0,1){30}}
\put(29,29){\line(0,1){30}}
\put(39,29){\line(0,1){30}}
\put(4.5,4){\circle*{3}}
\put(4.5,14){\circle*{3}}
\put(4.5,24){\circle*{3}}
\put(4.5,34){\circle*{3}}
\put(4.5,44){\circle*{3}}
\put(4.5,54){\circle*{3}}
\put(14.5,24){\circle*{3}}
\put(14.5,34){\circle*{3}}
\put(14.5,44){\circle*{3}}
\put(14.5,54){\circle*{3}}
\put(24.5,34){\circle*{3}}
\put(24.5,44){\circle*{3}}
\put(24.5,54){\circle*{3}}
\put(34.5,34){\circle*{3}}
\put(34.5,44){\circle*{3}}
\put(34.5,54){\circle*{3}}
\put(44.5,44){\circle*{3}}
\put(44.5,54){\circle*{3}}
\put(54.5,54){\circle*{3}}
\put(64.5,54){\circle*{3}}
\end{picture}
}
\end{picture}
\vspace{2ex} \\
\refstepcounter{figure}
Fig.\thefigure : An example of partition counted by $Q_{3,1}(7,6:20)$.
\addcontentsline{lof}{figure}%
{\protect\numberline{\thefigure}
{An example of partition counted by $Q_{3,1}(7,6:20)$.}}
\label{fig:(3,1)}
\end{center}
\end{minipage}

{}~

{\it Remark. } In a similar way given in
the above remark, we obtain
$$
{\cal Q}_{3,1}(n+1,n;q)= \prod_{j=1}^{n} (1+q^{2j}).
$$

{}~

\section{Proof of Main Propositions}

{\bf Proof of Proposition \ref{prop:k,i,1}. }
We prove Proposition \ref{prop:k,i,1} by induction.
We already showed the case $k=1$ by (\ref{eqn:1,1}),
and the case $k=2$ by Lemmas
\ref{lem:k=2,i=2,1} and \ref{lem:k=2,i=1,1}.
Now, fix a set $(k,i)$
such that $k\geq 3$, $2\leq i \leq k$, and
suppose that the claim of Proposition \ref{prop:k,i,1}
holds for $(k-1,i-1)$: i.e., for any $n', n_2 , N'$
there exists one-to-one correspondence
between
partition of $N'$ counted by
$Q_{2k-i,i-1}([\frac{n'+1+k-i}{2}],
[\frac{n'-k+i}{2}] ; n_2 ; N')$
and that of the form
\begin{equation}
N'=n_2^2 + \cdots + n_{k-1}^2 +n_i + \cdots + n_{k-1}
+\sum_{j=2}^{k-1} \sum_{p=1}^{n_j -n_{j+1}}
a^{(j)}_{p} ,
\label{eqn:hyp1}
\end{equation}
where
$$
n_2 \geq \cdots \geq n_{k-1}\geq n_k =0, ~~~~
2(n_2 + \cdots + n_{k-1}) \leq n'-k+i,
$$
and
\begin{equation}
\begin{array}{l}
n'-2(n_2 + \cdots + n_j )+
\alpha^{(k-1)}_{i-1 j-1}
\geq a^{(j)}_1 \geq \cdots \geq a^{(j)}_{n_j -n_{j+1}}
\geq 0, ~~~~2 \leq j \leq k-1.
\end{array}
\label{eqn:ind1}
\end{equation}

Fix one of graphs $G$ of partition of $N'$
with the maximal size
$([\frac{n'+1+k-i}{2}], [\frac{n'-k+i}{2}])$, with
$n_2$ right angles, and
all $n_2 $ successive ranks belong to
the interval $[-i+3,2k-i-2]$.
In what follows we consider two cases, separately.
Let us distinguish them, say, the Case A and the Case B.
For the Case A,
we wish to transform $G$ into $G^{(3)}$
of a partition of
\begin{equation}
N =N' +n_1 ^2 + a^{(1)}_1 + \cdots + a^{(1)}_{n_1 -n_2},
{}~~~~a^{(1)}_1 \geq \cdots \geq a^{(1)}_{n_1 -n_2} \geq 0,
\label{eqn:N',N;A,1}
\end{equation}
with the maximal size
$([\frac{n+1+k-i}{2}],[\frac{n-k+i}{2}])$,
with $n_1 (\geq n_2 )$ right angles,
and all $n_1 $ successive ranks
belong to the interval $[-i+2,2k-i-1]$, where $2 \leq i \leq k$.
When $i=2$,
we have another possibility: i.e.,
we can transform $G$ into a partition of
\begin{equation}
N =N' +n_1 (n_1 +1) + a^{(1)}_1 + \cdots + a^{(1)}_{n_1 -n_2},
{}~~~~a^{(1)}_1 \geq \cdots \geq a^{(1)}_{n_1 -n_2} \geq 0,
\label{eqn:N',N;B,1}
\end{equation}
with the maximal size
$([\frac{n+k}{2}],[\frac{n-k+1}{2}])$,
with $n_1 (\geq n_2 )$ right angles,
and all $n_1 $ successive ranks
belong to the interval $[1,2k-2]$.
This is the case B. (See {\it 2nd step} below in detail.)

We notice that now $n,n_1$ and $N$ are fixed
while $n',n_2$ and $N'$ should be regarded
as variables. In fact, we will show $n'=n-2n_1$
(resp. $n'=n-2n_1 -1$) for the Case A (resp. Case B),
later.

The transformation from $G$ to $G^{(3)}$
consists of three steps.
The first two steps will increase $N'$ by $n_1^2 $
(resp. $n_1 (n_1 +1)$) for the Case A (resp. Case B),
in the last step we will corporate $n_1 -n_2 $ non negative integers.

{\it 1st step. } In this step,
we transform $G$ to $G^{(1)}$ as follows.
If $r_1 (G)\geq k-i+1$ (resp. $\leq k-i$),
add $n_2$ (resp. $n_2 -1$) dots to the right of the first row,
and add $n_2 -1$ (resp. $n_2 $) dots below the first column.
If $r_2 (G) \geq k-i+1$ (resp. $\leq k-i$),
add $n_2 -1$ (resp. $n_2 -2$) dots to the right of the second row,
and add $n_2 -2$ (resp. $n_2 -1$) dots below the second column.
In general,
If $r_l (G) \geq k-i+1$ (resp. $\leq k-i$),
add $n_2 -(l-1)$ (resp. $n_2 -l$) dots to the right of the $l$th row,
and add $n_2 -l$ (resp. $n_2 -(l-1)$) dots below the $l$th column.
The graph thus obtained is $G^{(1)}$.
This is also a Ferrers graph, because by the construction
the number of dots on
the $l$th row (resp. column) of $G^{(1)}$ are not larger
than that of the $(l-1)$st row (resp. column).

Note that
the number of dots increase by $(2n_2 -1)+(2n_2 -3)+
\cdots + 1 =n_2^2 $,
and $-i+2 \leq r_l (G^{(1)}) \leq k-i-1$, or
$k-i+2 \leq r_l (G^{(1)}) \leq 2k-i-1$,
for $1 \leq l \leq n_2$.

Fig.\ref{fig:(4,1)} is an example for $(k-1,i-1)=(2,1)$.
In this case Fig.\ref{fig:1st} gives $G^{(1)}$.

\begin{minipage}[t]{\minitwocolumn}
\begin{center}
\ \vspace{2ex}\\
\setlength{\unitlength}{0.5mm}
\begin{picture}(100,60)
\put(-18,0){\begin{picture}(101,0)
\put(4.5,4){\circle*{3}}
\put(4.5,14){\circle*{3}}
\put(4.5,24){\circle*{3}}
\put(4.5,34){\circle*{3}}
\put(4.5,44){\circle*{3}}
\put(4.5,54){\circle*{3}}
\put(4.5,64){\circle*{3}}
\put(14.5,24){\circle*{3}}
\put(14.5,34){\circle*{3}}
\put(14.5,44){\circle*{3}}
\put(14.5,54){\circle*{3}}
\put(14.5,64){\circle*{3}}
\put(24.5,34){\circle*{3}}
\put(24.5,44){\circle*{3}}
\put(24.5,54){\circle*{3}}
\put(24.5,64){\circle*{3}}
\put(34.5,44){\circle*{3}}
\put(34.5,54){\circle*{3}}
\put(34.5,64){\circle*{3}}
\put(44.5,54){\circle*{3}}
\put(44.5,64){\circle*{3}}
\put(54.5,54){\circle*{3}}
\put(54.5,64){\circle*{3}}
\put(64.5,54){\circle*{3}}
\put(64.5,64){\circle*{3}}
\put(74.5,54){\circle*{3}}
\put(74.5,64){\circle*{3}}
\put(84.5,64){\circle*{3}}
\put(94.5,64){\circle*{3}}
\end{picture}
}
\end{picture}
\vspace{2ex} \\
\refstepcounter{figure}
Fig.\thefigure : The resulting graph after {\it 1st step}
obtained from Fig.\ref{fig:(4,1)}.
\addcontentsline{lof}{figure}%
{\protect\numberline{\thefigure}
{The resulting graph after {\it 1st step} obtained from
Fig.\ref{fig:(4,1)}.}}
\label{fig:1st}
\end{center}
\end{minipage}

{}~

{\it Remark. }
Suppose that
$-i+2 \leq r_l (G^{(1)}) \leq k-i-1$ or
$k-i+2 \leq r_l (G^{(1)}) \leq 2k-i-1$, for
$1 \leq l \leq n_2$.
Then
if $r_l (G^{(1)})\geq k-i+2$ (resp. $\leq k-i-1$),
remove $n_2 -(l-1)$ (resp. $n_2 -l$) dots
on the $l$th row,
and remove $n_2 -l$ (resp. $n_2 -(l-1)$)
dots on the $l$th column.
This is the original $G$.
Thus {\it 1st step} is invertible.

{\it 2nd step. }
{}From this step, we have to consider the Case A and B separately.

\underline{Case A. } Now we wish to add
additional $(n_1 -n_2) \times (n_1 -n_2 )$
square to this graph.
For that purpose,
add $n_1 -n_2 $ dots to the right of
the $l$th row and below the $l$th column, where $1 \leq l \leq n_2$.
Then we can add an $(n_1 -n_2) \times (n_1 -n_2 )$
square in the top-left corner
of the $n_2 $th proper subgraph.
This resulting graph is $G^{(2)}$.
Now the total increase of the number of dots is
$$
n_2^2 +2n_2 (n_1 -n_2 ) +(n_1 -n_2 )^2 =n_1 ^2.
$$
This accounts the additional $n_1^2 $ in (\ref{eqn:N',N;A,1}).

Note that at this stage
$-i+2 \leq r_l (G^{(2)}) \leq k-i-1$ or
$k-i+2 \leq r_l (G^{(2)}) \leq 2k-i-1$
for $1 \leq l \leq n_2 $, and
$r_l (G^{(2)})=0$ for $n_2 +1 \leq l \leq n_1 $.

Fig.\ref{fig:2nd,A} is an example of
$G^{(2)}$ obtained from Fig.\ref{fig:1st},
where we set $n_1 -n_2 =2$.

\underline{Case B. } When $i=2$,
we have two choices to add an
$(n_1 -n_2)\times (n_1 -n_2)$ square or an
$(n_1 -n_2) \times (n_1 -n_2 +1)$ rectangle.
The latter one corresponds to the Case B.
In this case
for $1 \leq l \leq n_2 $, we
add $n_1 -n_2 +1 $ dots to the right of the $l$th row
and
add $n_1 -n_2 $ dots below the $l$th column.
After that we add an
$(n_1 -n_2) \times (n_1 -n_2 +1)$ rectangle
at the $n_2 $th proper subgraph.
This resulting graph is $G^{(2)}$.
Now the total increase of the number of dots
is
$$
n_2^2 +n_2 (n_1 -n_2 )+
n_2 (n_1 -n_2 +1)+(n_1 -n_2)(n_1 -n_2 +1)=
n_1^2 +n_1.
$$
This accounts the additional $n_1 (n_1 +1)$ in (\ref{eqn:N',N;B,1}).

Note that at this stage
$1 \leq r_l (G^{(2)}) \leq k-2$ or
$k+1 \leq r_l (G^{(2)}) \leq  2k-2$
for $1 \leq l \leq n_2 $, and
$r_l (G^{(2)})=1$ for $n_2 +1 \leq l \leq n_1 $.

Fig.\ref{fig:2nd,B} represents $G^{(2)}$
obtained from Fig.\ref{fig:1st},
where we set $n_1 -n_2 =2$.

\begin{minipage}[t]{\minitwocolumn}
\begin{center}
\ \vspace{2ex}\\
\setlength{\unitlength}{0.5mm}
\begin{picture}(100,80)
\put(-18,0){\begin{picture}(101,0)
\put(4.5,4){\circle*{3}}
\put(4.5,14){\circle*{3}}
\put(4.5,24){\circle*{3}}
\put(4.5,34){\circle*{3}}
\put(4.5,44){\circle*{3}}
\put(4.5,54){\circle*{3}}
\put(4.5,64){\circle*{3}}
\put(4.5,74){\circle*{3}}
\put(4.5,84){\circle*{3}}
\put(14.5,24){\circle*{3}}
\put(14.5,34){\circle*{3}}
\put(14.5,44){\circle*{3}}
\put(14.5,54){\circle*{3}}
\put(14.5,64){\circle*{3}}
\put(14.5,74){\circle*{3}}
\put(14.5,84){\circle*{3}}
\put(24.5,34){\circle*{3}}
\put(24.5,44){\circle*{3}}
\put(24.5,54){\circle*{3}}
\put(24.5,64){\circle*{3}}
\put(24.5,74){\circle*{3}}
\put(24.5,84){\circle*{3}}
\put(34.5,44){\circle*{3}}
\put(34.5,54){\circle*{3}}
\put(34.5,64){\circle*{3}}
\put(34.5,74){\circle*{3}}
\put(34.5,84){\circle*{3}}
\put(44.5,44){\circle*{3}}
\put(44.5,54){\circle*{3}}
\put(44.5,64){\circle*{3}}
\put(44.5,74){\circle*{3}}
\put(44.5,84){\circle*{3}}
\put(54.5,64){\circle*{3}}
\put(54.5,74){\circle*{3}}
\put(54.5,84){\circle*{3}}
\put(64.5,74){\circle*{3}}
\put(64.5,84){\circle*{3}}
\put(74.5,74){\circle*{3}}
\put(74.5,84){\circle*{3}}
\put(84.5,74){\circle*{3}}
\put(84.5,84){\circle*{3}}
\put(94.5,74){\circle*{3}}
\put(94.5,84){\circle*{3}}
\put(104.5,84){\circle*{3}}
\put(114.5,84){\circle*{3}}
\end{picture}
}
\end{picture}
\vspace{2ex} \\
\refstepcounter{figure}
Fig.\thefigure : The resulting graph after 2nd step.
$n_1 -n_2 =2$. (Case A).
\addcontentsline{lof}{figure}%
{\protect\numberline{\thefigure}
{The resulting graph after 2nd step.
$n_1 -n_2 =2$. (Case A).}}
\label{fig:2nd,A}
\end{center}
\end{minipage}%
\hspace{\columnsep}%
\begin{minipage}[t]{\minitwocolumn}
\begin{center}
\ \vspace{2ex}\\
\setlength{\unitlength}{0.5mm}
\begin{picture}(100,80)
\put(-18,0){\begin{picture}(101,0)
\put(4.5,4){\circle*{3}}
\put(4.5,14){\circle*{3}}
\put(4.5,24){\circle*{3}}
\put(4.5,34){\circle*{3}}
\put(4.5,44){\circle*{3}}
\put(4.5,54){\circle*{3}}
\put(4.5,64){\circle*{3}}
\put(4.5,74){\circle*{3}}
\put(4.5,84){\circle*{3}}
\put(14.5,24){\circle*{3}}
\put(14.5,34){\circle*{3}}
\put(14.5,44){\circle*{3}}
\put(14.5,54){\circle*{3}}
\put(14.5,64){\circle*{3}}
\put(14.5,74){\circle*{3}}
\put(14.5,84){\circle*{3}}
\put(24.5,34){\circle*{3}}
\put(24.5,44){\circle*{3}}
\put(24.5,54){\circle*{3}}
\put(24.5,64){\circle*{3}}
\put(24.5,74){\circle*{3}}
\put(24.5,84){\circle*{3}}
\put(34.5,44){\circle*{3}}
\put(34.5,54){\circle*{3}}
\put(34.5,64){\circle*{3}}
\put(34.5,74){\circle*{3}}
\put(34.5,84){\circle*{3}}
\put(44.5,44){\circle*{3}}
\put(44.5,54){\circle*{3}}
\put(44.5,64){\circle*{3}}
\put(44.5,74){\circle*{3}}
\put(44.5,84){\circle*{3}}
\put(54.5,44){\circle*{3}}
\put(54.5,54){\circle*{3}}
\put(54.5,64){\circle*{3}}
\put(54.5,74){\circle*{3}}
\put(54.5,84){\circle*{3}}
\put(64.5,64){\circle*{3}}
\put(64.5,74){\circle*{3}}
\put(64.5,84){\circle*{3}}
\put(74.5,74){\circle*{3}}
\put(74.5,84){\circle*{3}}
\put(84.5,74){\circle*{3}}
\put(84.5,84){\circle*{3}}
\put(94.5,74){\circle*{3}}
\put(94.5,84){\circle*{3}}
\put(104.5,74){\circle*{3}}
\put(104.5,84){\circle*{3}}
\put(114.5,84){\circle*{3}}
\put(124.5,84){\circle*{3}}
\end{picture}
}
\end{picture}
\vspace{2ex} \\
\refstepcounter{figure}
Fig.\thefigure : The resulting graph after 2nd step.
$n_1 -n_2 =2$. (Case B).
\addcontentsline{lof}{figure}%
{\protect\numberline{\thefigure}
{The resulting graph after 2nd step.
$n_1 -n_2 =2$. (Case B).}}
\label{fig:2nd,B}
\end{center}
\end{minipage}

{}~

{\it Remark. }
By the construction,
$G^{(2)}$ has $n_1 $ right angles.
Since we added at least one dots
for the first $n_2 $ right angles
at {\it 1st step},
the smallest number $l$ such that
$l$th subgraph is an $(n_1 -l)\times (n_1 -l)$ square
(resp. $(n_1 -l)\times (n_1 -l+1)$ rectangle) is
equal to $n_2 $,
for the Case A (resp. Case B).
Remove $n_2 $th proper subgraph of $G^{(2)}$,
and then remove $n_1 -n_2 $ dots on each column and
$n_1 -n_2 $ (resp. $n_1 -n_2 +1$) dots on each row
for the Case A (resp. Case B).
Then we obtain $G^{(1)}$.
Thus {\it 2nd step} is invertible.

{\it 3rd step. } Now we are in a position to corporate $n_1 -n_2$
non negative parts
$a^{(1)}_1 \geq \cdots \geq a^{(1)}_{n_1 -n_2} \geq 0$.
For simplicity let us denote them as
$a_1 \geq \cdots \geq a_{n_1 -n_2} \geq 0$.
This step consists of $n_1 -n_2 $ substeps, each of which
corresponds to the procedure of adding $a_p$ dots,
where $1 \leq p \leq n_1 -n_2$.

In the argument presented below, $c$ denotes
$k-i$ (resp. $k-1$) for the Case A (resp. Case B).
For a Ferrers graph corresponding to a certain partition,
if the graph obtained by adding or removing
a dot is a still
Ferrers graph, then such a manipulation is called {\it admissible}.

{\it 1st substep. }
We add $a_1 $ dots to $G^{(2)}$ following the Rules.
First we set $l=n_2 +1$.

Rule I($l$).
If $r_l (G^{(2)}) <c$ (resp. $>c+1$),
we add dots to the right of the $l$th row
(resp. below the $l$th column)
whenever admissible,
until the total number of dots reaches $a_1$ or
the $l$th successive rank reaches $c$.
If we can add $a_1$ dots in this way,
we go to the next substep.

Rule II($l$). Starting from the $l$th row
we add a dot on the
$l$th row and column in turn, whenever admissible.
In other words,
we add a dot to the right of the $l$th row
if admissible.
Then we add a dot below
the $l$th column, if admissible. After that we add a dot
to the right of the $l$th row, if admissible, and so on.
If we can add $a_1$ dots in this way,
we go to the next substep.

Note that at this stage, the $l$th successive rank
takes the values $c+1$ and $c$, in turn.

Rule III($l$). When
$r_{l-1} (G^{(2)}) < c$ (resp. $> c+1$),
it may happen that
the manipulation of adding a dot to the right
of the $l$th row (resp. below
the $l$th column) is not admissible.
In such a case we add dots below
the $l$th column (resp. to the right of $l$th row)
whenever admissible. If we can add $a_1$ dots in this way,
we go to the next substep.

Unfortunately, however, maybe we cannot add a dot
below the $l$th column
(resp. to the right of the $l$th row)
any more without breaking the admissibility,
before finishing to add $a_1$ dots.
Then we reset $l=n_2 $
and repeat the above manipulations.
In general when we cannot add a dot any more
following the Rule III($l$),
without breaking the admissibility,
we replace $l$ by $l-1$ and repeat the above manipulations.
In this way, eventually we can add $a_1 $ dots.
Note that
we will have no chance to follow the Rule III(1).

The following figure gives the way how to add
$a_1 (=16)$ dots to Fig.\ref{fig:2nd,A}.

\begin{minipage}[t]{\minitwocolumn}
\begin{center}
\ \vspace{2ex}\\
\setlength{\unitlength}{0.5mm}
\begin{picture}(140,110)
\put(8,30){\begin{picture}(101,0)
\put(4.5,4){\circle*{3}}
\put(4.5,14){\circle*{3}}
\put(4.5,24){\circle*{3}}
\put(4.5,34){\circle*{3}}
\put(4.5,44){\circle*{3}}
\put(4.5,54){\circle*{3}}
\put(4.5,64){\circle*{3}}
\put(4.5,74){\circle*{3}}
\put(4.5,84){\circle*{3}}
\put(14.5,24){\circle*{3}}
\put(14.5,34){\circle*{3}}
\put(14.5,44){\circle*{3}}
\put(14.5,54){\circle*{3}}
\put(14.5,64){\circle*{3}}
\put(14.5,74){\circle*{3}}
\put(14.5,84){\circle*{3}}
\put(24.5,34){\circle*{3}}
\put(24.5,44){\circle*{3}}
\put(24.5,54){\circle*{3}}
\put(24.5,64){\circle*{3}}
\put(24.5,74){\circle*{3}}
\put(24.5,84){\circle*{3}}
\put(34.5,44){\circle*{3}}
\put(34.5,54){\circle*{3}}
\put(34.5,64){\circle*{3}}
\put(34.5,74){\circle*{3}}
\put(34.5,84){\circle*{3}}
\put(44.5,44){\circle*{3}}
\put(44.5,54){\circle*{3}}
\put(44.5,64){\circle*{3}}
\put(44.5,74){\circle*{3}}
\put(44.5,84){\circle*{3}}
\put(54.5,64){\circle*{3}}
\put(54.5,74){\circle*{3}}
\put(54.5,84){\circle*{3}}
\put(64.5,74){\circle*{3}}
\put(64.5,84){\circle*{3}}
\put(74.5,74){\circle*{3}}
\put(74.5,84){\circle*{3}}
\put(84.5,74){\circle*{3}}
\put(84.5,84){\circle*{3}}
\put(94.5,74){\circle*{3}}
\put(94.5,84){\circle*{3}}
\put(104.5,84){\circle*{3}}
\put(114.5,84){\circle*{3}}
\put(53.0,52){$1$}
\put(33.0,32){$2$}
\put(63.0,62){$3$}
\put(73.0,62){$4$}
\put(23.0,22){$5$}
\put(83.0,62){$6$}
\put(93.0,62){$7$}
\put(13.0,12){$8$}
\put(13.0,2){$9$}
\put(101.0,72){$10$}
\put(111.0,72){$11$}
\put(0.5,-8){$12$}
\put(0.5,-18){$13$}
\put(120.5,82){$14$}
\put(0.5,-28){$15$}
\put(130.5,82){$16$}
\end{picture}
}
\end{picture}
\vspace{2ex} \\
\refstepcounter{figure}
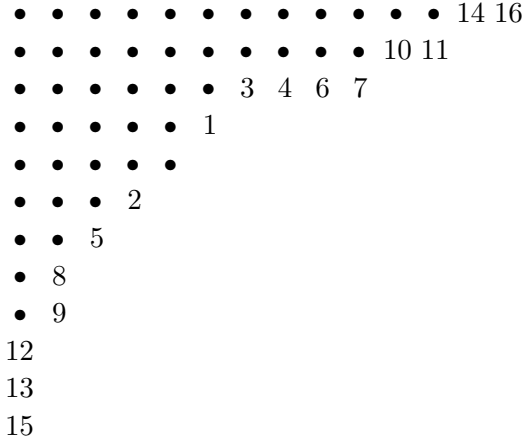
Fig.\thefigure : The order of addition of $a_1 (=16)$ dots.
\label{fig:3rd}
\end{center}
\end{minipage}%

{}~

{\it 2nd substep. }
The aim of this substep is to add $a_2$ dots.
Replace $n_2 $ by $n_2 +1$ and $a_1 $ by $a_2$.
Then repeat the above procedure.
Note that we can add up to $a_1 $
by the construction.
Graphically speaking, $p$th added dot in
this substep locates below or to the right of
$p$th added one in the last substep.

{\it $p$th substep. } In general,
after {\it $(p-1)$st substep},
we repeat this procedure
under the replacement of $n_2 $ by $n_2 +1$
and $a_{p-1}$ by $a_p$.

It is evident from the construction that
all the successive ranks of
the resulting graph $G^{(3)}$
lie in the interval $[-i+2, 2k-i-1]$
(resp. $[1, 2k-2]$) for
the Case A (resp. Case B).

Now we would like to consider
the maximal size of the original graph $G$
and the maximal value of $a_1$
for fixed $k,i,n_1 $ and $n$.

\underline{Case A. }
Let $\mu_1 $ and $\nu_1 $ be the length of
the first row and column of
the original graph $G$, respectively.
In {\it 1st step}, these length increase
by $n_2 $ and $n_2 -1$ (resp. $n_2 -1$ and $n_2 $)
if $\mu_1 -\nu_1 \geq c+1$ (resp. $\mu_1 -\nu_1 \leq c$).
In {\it 2nd step}, they both increase
by $n_1 -n_2 $.
Thus after {\it 2nd step}
the length of the first row and column
increase by $n_1 $ and $n_1 -1$
(resp. $n_1 -1$ and $n_1 $)
if $\mu_1 -\nu_1 \geq c+1$ (resp. $\mu_1 -\nu_1 \leq c$).
Hence we have
$\mu_1 +n_1 \leq [\frac{n+1+k-i}{2}]$ and
$\nu_1 +n_1 \leq [\frac{n-k+i}{2}]$.
The maximal values of $\mu_1 $ and $\nu_1 $
are $[\frac{n'+1+k-i}{2}]$ and $[\frac{n'-k+i}{2}]$,
respectively. In other words,
$n'$ is the possible maximal number that satisfies
both
$[\frac{n'+1+k-i}{2}] +n_1 \leq [\frac{n+1+k-i}{2}]$ and
$[\frac{n'-k+i}{2}] +n_1 \leq [\frac{n-k+i}{2}]$.
Therefore we conclude that $n' =n-2n_1 $.
Consequently (\ref{eqn:ind1}) reads as
\begin{equation}
\begin{array}{l}
n-2n_1 -2(n_2 + \cdots + n_j )-
\alpha^{(k-1)}_{i-1 j-1}
\geq a^{(j)}_1 \geq \cdots \geq a^{(j)}_{n_j -n_{j+1}}
\geq 0,
\end{array}
\label{eqn:indafter1}
\end{equation}
which implies $\alpha^{(k)}_{ij}=\alpha^{(k-1)}_{i-1 j-1}$
for $2 \leq i \leq k, 2 \leq j \leq k-1$.

Next let us determine the maximal value of $a_1 $.
Every time we add a dot, we
project each dot to the first row
(resp. column) orthogonally if we added it
on the $l$th row (resp. column)
following one of the Rules I($l$), II($l$) and
III($l$).
The lattice site $(l_1 , l_2 )$ refers to
the intersection point of $l_1 $th column
and $l_2 $th row.
Following one of the Rule I($l$), II($l$) and III($l$),
we can put a dot at $(m,l)$
(resp. $(l,m)$) if and only if $(m,l)$ (resp. $(l,m)$)
is not occupied yet,
all $(l_1 , l)$ with $l_1 < m$
(resp. $(l, l_2)$ with $l_2 < m$)
are already occupied, and $m \leq r$,
where $r$ is the number of dots on
the $(l-1)$st row (resp. column).
Thus $m>n_1 $ because $G^{(2)}$ has an $n_1 \times n_1 $
square in the top-left corner.
When we put dots on the $(l-1)$st row (resp. column)
following one of the Rules I($l-1$), II($l-1$) and III($l-1$),
we begin by adding a dot at $(r+1,l-1)$ (resp. $(l-1,r+1)$).
Consequently, projected dots occupy the first row (resp. column)
starting from $(n_1 +1, 1)$ (resp. $(1, n_1 +1)$)
to outside. (See Fig.\ref{fig:3rd}.)
This observation immediately derive
that
\begin{equation}
n-2n_1 \geq a_1 \geq \cdots \geq a_{n_1 -n_2} \geq 0,
\label{eqn:max,1}
\end{equation}
which implies $\alpha^{(k)}_{i1}=0$ for
$2 \leq i \leq k$.
Two relations (\ref{eqn:indafter1}--\ref{eqn:max,1})
reproduce the finiteness condition (\ref{eqn:fincond1})
with (\ref{eqn:alpha}) for $2 \leq i \leq k$.

\underline{Case B. }
Let $\mu_1 $ and $\nu_1 $ be
the length of the first row and column, respectively.
By the parallel argument
above, after {\it 2nd step}
the length of the first row and column
increase by $n_1 +1$ and $n_1 -1$ (resp.
$n_1 $ and $n_1 $) if $\mu_1 -\nu_1 \geq k-1$
(resp. $\mu_1 -\nu_1 \leq k-2$).
In this case, $n'$ is the maximal number that
satisfies both
$[\frac{n'+k-1}{2}] +n_1 +1 \leq [\frac{n+k}{2}]$ and
$[\frac{n'-k+2}{2}] +n_1 \leq [\frac{n-k+1}{2}]$.
Therefore we conclude that $n'=n -2n_1 -1$.
Consequently (\ref{eqn:ind1}) becomes
\begin{equation}
\begin{array}{l}
n-2n_1 -1-2(n_2 + \cdots + n_j )-
\alpha^{(k-1)}_{1 j-1}
\geq a^{(j)}_1 \geq \cdots \geq a^{(j)}_{n_j -n_{j+1}}
\geq 0,
\end{array}
\label{eqn:indafter1,1}
\end{equation}
which implies $\alpha^{(k)}_{1 j}=\alpha^{(k-1)}_{1 j-1}+1$ for
$2 \leq j \leq k-1$.
As for the maximal value of $a_1 $,
from the parallel argument above, we obtain
\begin{equation}
n-(2n_1 +1)\geq a_1 \geq \cdots \geq a_{n_1 -n_2} \geq 0,
\label{eqn:maxB,1}
\end{equation}
which implies $\alpha^{(k)}_{1 1}=1$.
Two relations (\ref{eqn:indafter1,1}--\ref{eqn:maxB,1})
reproduce the finiteness condition (\ref{eqn:fincond1})
with (\ref{eqn:alpha})
for $i=1$.

Thus the remaining work is to show that
for any graph $G^{(3)}$
with the maximal size $([\frac{n+1+k-i}{2}],
[\frac{n-k+i}{2}]$), with $n_1 $ right angles, and
all $n_1 $ successive ranks
lie in the interval $[-i+2, 2k-i-1]$,
we can extract information of
$n_1 -n_2 $ and
$a_1 \geq \cdots \geq a_{n_1 -n_2 } \geq 0$
by reversing {\it 3rd step}.
Because by taking into account remarks at the end of
{\it 1st step} and {\it 2nd step},
we can reconstruct $G$, which
reduces to the assumption of the induction.

Suppose that we add a dot on the $l$th
right angle
following the Rule II($l$),
the successive rank of the $l$th right angle
takes the values $c+1$ and $c$, in turn. This will break
when the length of the $l$th shorter side relative to $c$
reaches that of the $(l-1)$st shorter side relative to $c$.
In that case,
we have to begin to add dots
on the $l$th right angle following the Rule III($l$).
When we add dots on the $(l-1)$th right angle
following the Rule I($l-1$), the
$(l-1)$st and the $l$th longer side relative to $c$
have the common length and direction,
and the $(l-1)$st shorter side relative to $c$
is longer than the $l$th one.
Special consideration is required
when we begin to add dots
following the Rule I($n_2 +p$) in {\it $p$th substep}.
In this case we always add dots to the right of the $(n_2 +p)$th row
because $r_{n_2 +p}(G^{(2)})=0$ (resp. $=1$)
$\leq c$ for the Case A (resp. Case B), and thus
the length of the $(n_2 +p)$th column
is equal to a given number $n_1 $.

Therefore
we conclude that
the last added dot among $a_p$ ($1 \leq p \leq n_1 -n_2$)
dots is located at the $l$th right angle,
if $l$ satisfies
at least one of the following four conditions:

(1) $r_l (G^{(3)}) =c$, or $c+1$;

(2) The $l$th shorter side relative to $c$
has the same length and direction as
those of the $(l-1)$st shorter side relative to $c$;

(3) The $l$th longer side relative to $c$ has
the same length and direction as those of
the $(l+1)$st longer side relative to $c$,
and the $l$th shorter side relative to $c$
is longer than the $(l+1)$st shorter side relative to $c$;

(4) The length of the $l$th column
is equal to $n_1 $.

{}From this observation,
we can reverse {\it 3rd step} as follows.
For any given graph $G^{(3)}$
counted by
$Q_{2k+1-i,i}([\frac{n+1+k-i}{2}],
              [\frac{n-k+i}{2}]; n_1 ; N)$,
we scan right angles from the most outer one.
Whenever we encounter a right angle which satisfies
at least one of the above four conditions, then
we mark that right angle.
The first marked right angle signifies the position
of the last added dot of $a_1 $ dots.
The second marked right angle signifies the position
of the last added dot of $a_2 $ dots.
In general,
The $p$th marked right angle signifies the position
of the last added dot of $a_p $ dots.
The total number of right angles we marked
is equal to $n_1 -n_2$.

Let us denote the $p$th marked
right angle by the $l_p$th right angle.
We begin by removing $a_{n_1 -n_2}$ dots.
Set $p=n_1 -n_2$.
If the $l_p$th successive rank is less than
$c$ (resp. larger than $c+1$), then
we remove dots on $l_p$th column
(resp. row)
until the $l_p$th
successive rank reaches $c$, whenever admissible.
This is the reverse manipulation of the Rule III($l_p$).
After that
we remove a dot
on the $l_p$th column and row in turn, such that
the $l_p$th successive rank takes values $c+1$ and $c$,
in turn. This is the reverse manipulation of
the Rule II($l_p$).
In the case we cannot remove a dot on the $l_p$th row
(resp. column)
any more without breaking the admissibility,
we remove dots on the $l_p$th column (resp. row)
whenever admissible.
This is the reverse manipulation of the Rule I($l_p$).
When
we cannot remove a dot on the $l_p$th column (resp. row)
any more without breaking the admissibility,
we repeat the above manipulations
by replacing $l_p $ by $l_p +1$.
We continue this procedure until we remove
all dots on the $(n_ 2 +p)$th right angle except the ones
within the $n_1 \times n_1 $ square
(resp. $n_1 \times (n_1 +1)$ rectangle) in the top-left corner
for the Case A (resp. Case B).
The total number of dots we remove is equal to $a_{p}$.
After that we repeat this procedure
under the replacement of $p$ by $p-1$.
In this way, we can determine
$0 \leq a_{n_1 -n_2} \leq \cdots \leq a_1$.
Thus this step is invertible.

Therefore, the claim of this Proposition was established.
{}~~~$\Box$

We obtain Theorem \ref{thm:poly1} as a Corollary
of Propositions \ref{prop:LHS1} and \ref{prop:k,i,1}.

{}~

{\bf Proof of Proposition \ref{prop:k,i,2}. }
We already showed $k=2$ by
Lemmas \ref{lem:k=2,i=2,2} and \ref{lem:k=2,i=1,2}.
Fix a set $(k,i)$ such that
$k \geq 3, 2 \leq i \leq k$.
Suppose that
for any non negative integers $n', n_2, N'$,
there exists a one-to-one correspondence
between partition of $N'$ counted by
$Q_{2k-i-1,i-1}(n'+k-i , n' ; n_2 ; N')$
and that of the form
\begin{equation}
N'=n_2^2 + \cdots + n_{k-1}^2 +n_i + \cdots + n_{k-1}
+\sum_{j=2}^{k-2} \sum_{p=1}^{n_j -n_{j+1}}
a^{(j)}_{p} +2 \sum_{p=1}^{n_{k-1}} b_p ,
\label{eqn:hyp2}
\end{equation}
where
$$
n_2 \geq \cdots \geq n_{k-1}\geq 0, ~~~~
n_2 + \cdots + n_{k-1} \leq n',
$$
and
\begin{equation}
\begin{array}{l}
2n'-2(n_2 + \cdots + n_j )+
\beta^{(k-1)}_{i-1 j-1}
\geq a^{(j)}_1 \geq \cdots \geq a^{(j)}_{n_j -n_{j+1}}
\geq 0, ~~~~2 \leq j \leq k-2, \\
n'-(n_2 + \cdots +n_{k-1})
\geq b_1 \geq \cdots \geq b_{n_{k-1}} \geq 0.
\end{array}
\label{eqn:ind2}
\end{equation}

We can prove this Proposition
in a perfectly parallel way
as we proved Proposition \ref{prop:k,i,1}.
The transformation rules for $(k,i)$ in the present
case is
exactly the same as the one for $(k,i)$ given
in the proof
of Proposition \ref{prop:k,i,1}.
The only difference is
the evaluation of the finiteness conditions.

Here we also discuss them
for the Case A and the Case B, separately.

\underline{Case A. }
By the parallel argument to before,
after {\it 2nd step}
the length of the first row and column
increase by $n_1 $ and $n_1 -1$
(resp. $n_1 -1$ and $n_1 $)
if the first successive rank is larger
or equal to $c+1$ (resp. less than or equal to $c$).
Hence $n'$ is the maximal value
that satisfies both $n' +k-i+n_1 \leq n+k-i$ and
$n' +n_1 \leq n$. Therefore
we conclude that $n'=n-n_1 $.
Consequently (\ref{eqn:ind2}) reads as
\begin{equation}
\begin{array}{l}
2(n-n_1 )-2(n_2 + \cdots + n_j )+
\beta^{(k-1)}_{i-1 j-1}
\geq a^{(j)}_1 \geq \cdots \geq a^{(j)}_{n_j -n_{j+1}}
\geq 0, \\
n-n_1 -(n_2 + \cdots +n_{k-1})
\geq b_1 \geq \cdots \geq b_{n_{k-1}} \geq 0,
\end{array}
\label{eqn:indafter2}
\end{equation}
which implies $\beta ^{(k)}_{ij}=\beta ^{(k-1)}_{i-1 j-1}$,
for $2 \leq i \leq k, 2 \leq j \leq k-2$.

As for the maximal value of $a_1 $, from
the parallel argument to before we have
\begin{equation}
n+(n+k-i)-2n_1 \geq a_1 \geq \cdots \geq a_{n_1 -n_2} \geq 0,
\label{eqn:max,2}
\end{equation}
which implies $\beta ^{(k)}_{i1}=k-i$.
Two relations (\ref{eqn:indafter2}--\ref{eqn:max,2})
reproduce the finiteness condition (\ref{eqn:fincond2})
with (\ref{eqn:beta}) for $2 \leq i \leq k$.

\underline{Case B. }
By the parallel argument
for the Case A, we again obtain $n=n'+n_1 $.
Consequently (\ref{eqn:ind2}) becomes
(\ref{eqn:indafter2}),
which implies $\beta^{(k)}_{1 j}=\beta^{(k-1)}_{1 j-1}$.

As for the maximal value of $a_1 $,
from the parallel argument above, we obtain
\begin{equation}
n+(n+k-1)-(2n_1 +1)\geq a_1 \geq \cdots \geq a_{n_1 -n_2} \geq 0,
\label{eqn:maxB,2}
\end{equation}
which implies $\beta^{(k)}_{11}=k-2$.
Two relations (\ref{eqn:indafter2},\ref{eqn:maxB,2})
reproduce the finiteness condition (\ref{eqn:fincond2})
with (\ref{eqn:beta})
for $i=1$.

Therefore, the claim of this Proposition was established.
{}~~~~$\Box$

We obtain Theorem \ref{thm:poly2} as a Corollary
of Propositions
\ref{prop:LHS2} and \ref{prop:k,i,2}.

{}~

\section{Discussion}

Some groups \cite{NRT,KNS,SB1} found
expressions of Virasoro characters in terms
of fermionic sum representation,
by using Bethe ansatz.
As a byproduct, they obtained the Rogers--Ramanujan type
identities including conjectures.
Such intimate
connection between physics and the Rogers--Ramanujan
type identities is really worth surprising.
As was mentioned in Introduction,
our original aim is to prove several conjectures
appeared in \cite{SB1}.
We will discuss this matter
and wish to prove those {\it mathematically}
in a separate paper.

We obtain the graphical proofs for
Propositions \ref{prop:k,i,1} and \ref{prop:k,i,2} by
translating Burge's correspondence \cite{Bur,AB,Bre4}
into language of partitions.
Burge's interpretation
for the multiple sums on the
RHS's of (\ref{eqn:Gor},\ref{eqn:GE})
reminds us the space of states
of the CTM Hamiltonian of
the generalized Hard Hexagon model \cite{KAW,BA}.
It is also interesting to study
relation among our graphical method,
Burge's correspondence, the theory
of the crystal base \cite{K}, etc.

We wish to add a few words to conclude the
present paper.
Lemma \ref{lem:exp} was one of key lemmas
to evaluate the LHS's of the polynomial identities.
This lemma
implies that for $\mu -\nu \leq a-1$
\begin{equation}
p_{a,b}(\mu , \nu ; \lambda ; N) =
p(\mu -\sum_{i=1}^{\lambda } a_i ,
  \nu +\sum_{i=1}^{\lambda } a_i ;
  N -\sum_{i=1}^{\lambda } (2i-1) a_i ),
\label{eqn:pp}
\end{equation}
where
$a_i =a$ (resp. $b$) when $i$ is odd (resp. even).
We notice that
the RHS of (\ref{eqn:pp}) for any $\mu , \nu$ satisfies
exactly the same recursion relation as
the top half of (\ref{eqn:recp}).

Let us introduce the integrated partition functions
$$
\begin{array}{rcl}
p(\mu ; N) & = & \displaystyle\sum_{\nu \geq 0} p(\mu , \nu ; N), \\
p_{a,b}(\mu ; \lambda ; N) & = &
\displaystyle\sum_{\nu \geq 0} p_{a,b}(\mu , \nu ; \lambda ; N).
\end{array}
$$
Bressoud's map given in \cite{Bre}
establishes a graphical one-to-one correspondence
between partitions counted by
$p_{a,b}(\mu ; \lambda ; N)$ and those counted by
$p(\mu -\sum_{i=1}^{\lambda } a_i ;
   N -\sum_{i=1}^{\lambda } (2i-1) a_i )$.
Unfortunately, his map does not give a one-to-one correspondence
between partitions counted by
$p_{a,b}(\mu , \nu ; \lambda ; N)$ and those counted by
$p(\mu -\sum_{i=1}^{\lambda } a_i ,
   \nu +\sum_{i=1}^{\lambda } a_i ;
   N -\sum_{i=1}^{\lambda } (2i-1) a_i )$,
under the condition $\mu -\nu \leq a-1$.
Thus it is still an open problem to show (\ref{eqn:pp})
graphically.

{}~

\section*{Acknowledgment}

We would like to thank Professors G. E. Andrews,
A. Kato and B. M. McCoy for useful information and
their interest in this work.

\section*{Note Added}

After finishing this work, G. E. Andrews
informed us that Theorem 1 in \cite{ABBBFV}
includes Propositions \ref{prop:LHS1} and \ref{prop:LHS2}
as special cases.
We have also received \cite{K} that containes an
independent proof of Theorem \ref{thm:poly1}.

\end{document}